\documentstyle[emulateapj,psfig,apjfonts]{article}
 
\def\gs{\mathrel{\raise0.35ex\hbox{$\scriptstyle >$}\kern-0.6em
\lower0.40ex\hbox{{$\scriptstyle \sim$}}}}
\def\ls{\mathrel{\raise0.35ex\hbox{$\scriptstyle <$}\kern-0.6em
\lower0.40ex\hbox{{$\scriptstyle \sim$}}}}

\makeatother

\lefthead{Smail et al.}
\righthead{Restframe Optical Properties of SMGs}
\submitted{Astrophysical Journal, 616, 71--85, 2004 November 20}

\begin{document}

\title
{The Rest-frame Optical Properties of SCUBA Galaxies}

\author{
Ian Smail,$^{\! 1}$ S.\,C.\ Chapman,$^{\! 2}$
A.\,W.\ Blain$^2$ \&
R.\,J.\ Ivison$^{3,4}$}

\altaffiltext{1}{Institute for Computational Cosmology, University of
  Durham, South Road,
        Durham DH1 3LE UK}
\altaffiltext{2}{California Institute of Technology, Pasadena, CA\,91125
        USA}
\altaffiltext{3}{Astronomy Technology Centre, Royal Observatory, 
        Blackford Hill, Edinburgh EH9 3HJ UK}
\altaffiltext{4}{Institute for Astronomy, University of Edinburgh, Royal
Observatory,  
        Blackford Hill, Edinburgh EH9 3HJ UK}

\setcounter{footnote}{4}
\slugcomment{Received 2004 May 28; accepted 2004 August 5}

\begin{abstract} We present optical and near-infrared photometry for a
sample of 96 dusty, far-infrared luminous galaxies.  We have precise
spectroscopic redshifts for all these galaxies yielding a median redshift
of $<\!z\!>=2.2$.  The majority, 78, are submillimeter-detected galaxies
lying at $z=0.2$--3.6, while the remaining 18 are optically-faint $\mu$Jy
radio galaxies at $z=0.9$--3.4 which are proposed to be similarly
luminous, dusty galaxies whose dust emission is too hot to be detected in
the submillimeter waveband.  We compare the photometric and morphological
properties of these distant, ultraluminous galaxies to local samples of
dusty, luminous galaxies.  We confirm that spectroscopically-identified
far-infrared luminous galaxies at $z>1$ display a wide variety in their
optical-near-infrared and near-infrared colors, with only a modest
proportion red enough to classify as unusually red. We show that on
average luminous, high-redshift dusty galaxies are both brighter and
redder in restframe optical passbands than comparable samples of
UV-selected star forming galaxies at similar redshifts.  Archival {\it
HST} ACS imaging of 20 of our galaxies demonstrates both morphological
indications of mergers and interactions, which may have triggered their
luminous far-infrared activity, and structured dust distributions within
these galaxies.  We derive a near-infrared Hubble diagram for far-infrared
luminous galaxies. This shows that this population is typically fainter
than high luminosity radio galaxies at similar redshifts and exhibit
significantly more scatter in their $K$-band magnitudes.  The restframe
optical luminosities of the far-infrared luminous population are
comparable to those of local ultraluminous infrared galaxies, although
their far-infrared luminosities are several times higher.  The typical
extinction-corrected optical luminosity of the high-redshift population,
assuming passive evolution, provides a good fit to the bright end of the
luminosity function of luminous spheroidal galaxies seen in rich clusters
at intermediate redshifts.  This adds to the growing body of evidence
showing that these high redshift, far-infrared luminous sources identify
star-formation and AGN fueling events in the early life of massive
galaxies in the Universe. 
\end{abstract}


\section{Introduction}

%
%
\setcounter{figure}{0}
\begin{figure*}
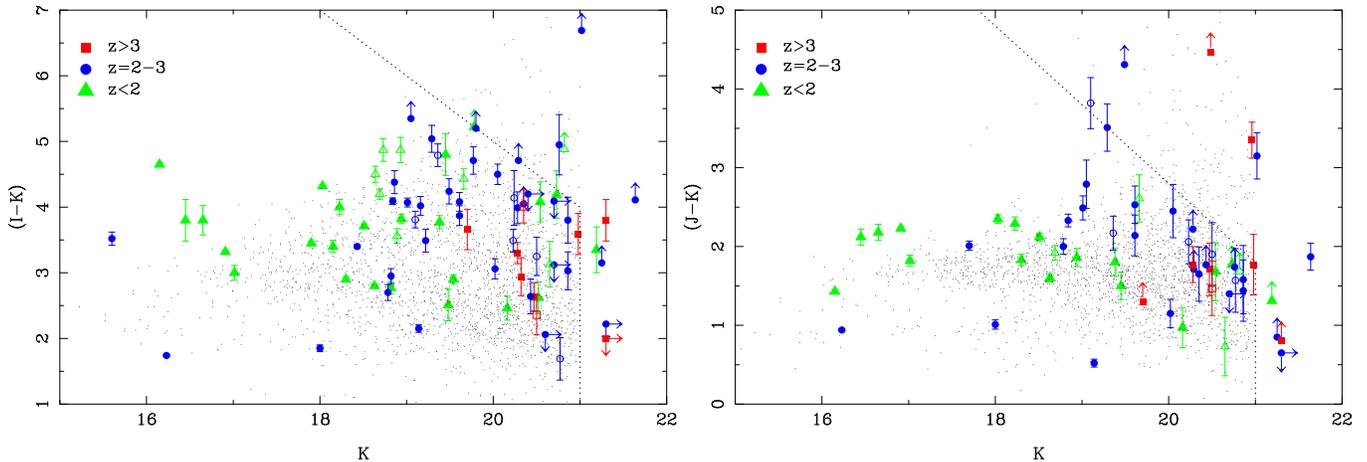

\centerline{\psfig{file=f1a.ps,width=3.5in,angle=0}
\psfig{file=f1b.ps,width=3.5in,angle=0}}
\caption{\small The $(I-K)$--$K$ and $(J-K)$--$K$ color-magnitude
diagrams for our sample of SMGs and OFRGs.  The SMGs are plotted as
filled symbols and the OFRGs are plotted as open symbols, with all
points coded on the galaxy redshifts.  We also indicate representative
3-$\sigma$ selection boundaries for the typical depth of our imaging.
For comparison we also show the distribution of observed colors of the
general population for a $K$-selected sample of galaxies in a
112 sq.\ arcmin region in the GOODS-North field (Bundy et al.\ 2004).
On average the SMG/OFRG sample are redder than the typical field
galaxies at their apparent magnitude: for example at $K=19$--20, the
SMG/OFRG have median colors of $(I-K)=4.07\pm 0.24$ and $(J-K)=2.49\pm
0.33$ whereas the general field colors are $(I-K)=2.82\pm 0.05$ and
$(J-K)=1.64\pm 0.02$.}
\end{figure*}

Ten billion years ago, at a redshift of $z\sim 2.5$, the Universe was a
much more active place. This is the era of the peak activity in quasars
(Boyle et al.\ 2000) and recent surveys suggest it coincides with the peak
activity in other galaxy populations.  These include surveys for star
forming galaxies selected either photometrically in their restframe UV
continuum light (Shapley et al.\ 2001; Steidel et al.\ 2004) or in their
restframe optical continuum light (Franx et al.\ 2003; Daddi et al.\ 2003,
2004; Cimatti et al.\ 2004), or through their intense far-infrared/radio
emission (Chapman et al.\ 2003a).  Taken together, the results from these
surveys provide a well-stocked, rich and apparently disparate zoo of
galaxies and AGN at this epoch. This enables us to investigate the
relationships between these different populations through direct
comparison of their physical properties, such as space densities, masses,
clustering strength, stellar ages and luminosities.  In this way we can
test whether these different classes represent, for example, a mass
sequence, or whether they are different stages in the life of a single
population, similar to the unification schemes which have been suggested
to connect local ultraluminous infrared galaxies (ULIRGs) and QSOs
(Sanders et al.\ 1988).

This paper focuses on determining the photometric properties in the
restframe optical wavebands of a large sample of far-infrared luminous
galaxies at $z\sim 2.5$ with complete redshift coverage (Chapman et al.\
2003a, 2004a, 2004b). These high-redshift, far-infrared luminous galaxies
are selected using two independent selection criteria.  The larger
subsample is based on a spectroscopic survey for optically-faint,
radio-selected sources with submillimeter (submm) fluxes above
$\sim$\,5\,mJy (Chapman et al.\ 2003a, 2004a). These are ``classical''
submm (or SCUBA) galaxies -- SMGs -- which appear to be a population of
highly obscured starbursts and starburst/AGN at high redshifts (Bertoldi
et al.\ 2000; Smail et al.\ 2002b; Ivison et al.\ 2002;  Webb et al.\
2003; Chapman et al.\ 2003a, 2004a;  Borys et al.\ 2004; Dannerbauer et
al.\ 2004) and are thought to be intimately linked with the formation of
massive galaxies (Frayer et al.\ 1998, 1999, 2004; Gear et al.\ 2000; Lutz
et al.\ 2001; Genzel et al.\ 2003; Blain et al.\ 2004a; Greve et al.\
2004a). The second, smaller, subsample are similarly optically-faint,
$\mu$Jy radio sources (OFRGs) with spectroscopic redshifts which place
them at comparable redshifts to the SMGs, but with limits on their submm
flux which preclude them being luminous submm sources (Chapman et al.\
2004b).  Although they are undetected in the submm waveband, these
galaxies are still proposed to be luminous far-infrared sources, but with
higher characteristic dust temperatures than those exhibited by the
submm-detected population (Blain et al.\ 2004b; Chapman et al.\ 2004b).  
This proposal will be tested through measurement of the mid-infrared
properties of these galaxies with the {\it Spitzer Space Telescope
(Spitzer)} in the near future.  Even if this claim proves to be erroneous
their inclusion in our analysis does not qualitatively change any of the
conclusions in this paper.

The selection function for this far-infrared luminous sample is very broad
and so to relate their properties to those of the populations of $z\sim
2.5$ galaxies and AGN selected via different techniques, we must have
redshift information. With this we can isolate those galaxies in the
relevant redshift range and simply compare their observed properties to
other populations at these redshifts, including BX/BM galaxies (Steidel et
al.\ 2004), very red near-infrared galaxies at $z\sim 2$ (Franx et al.\
2003) or QSOs (Boyle et al.\ 2000).  We are particularly interested in
testing the relationship between these populations by comparing their
stellar masses, as well as testing the possible connection between
SMG/OFRG and the formation phase of the luminous, evolved elliptical
galaxies which populate high-density regions out to at least $z\sim 1$.  
For galaxy populations at $z\sim 2$ these tests can be done using
characteristics derived from the restframe optical properties of the
galaxies -- which are accessible in the near-infrared wavebands from the
ground. Further progress will come from space-based, mid-infrared
observations with {\it Spitzer}.  These have the ability to probe the
restframe near-infrared emission from these galaxies, providing an even
more reliable measure of the luminosities of their stellar populations.

In this paper we collect optical and near-infrared photometry for a sample
of 96 submm- or radio-selected, far-infrared luminous galaxies with
precise spectroscopic redshifts.  We use these data to discuss the
restframe optical continuum properties of these galaxies and compare them
to similar observations of both low- and high-redshift galaxy populations.
We use our near-infrared photometry to estimate the likely optical
luminosities and model their colors to derive a typical age and dust
extinction for the population as a whole. We then examine the evolution of
these galaxies to determine the properties of possible descendents at
lower redshifts.

The paper is structured as follows: \S2 details our observations and their
reduction, while \S3 analyses these and gives the results derived from
these data.  \S4 discusses these results and summarises our main
conclusions. We assume through out a cosmology with $h_{100}=0.71$,
$\Omega_0=0.27$ and $\Omega_\Lambda=0.73$.

\section{Observations and Reduction}

Our analysis uses new near-infrared imaging in conjunction with archival
observations and published optical and near-infrared photometry to produce
a catalog of optical/near-infrared colors for spectroscopically-identified
submm and radio-selected sources from the surveys of Chapman et al.\
(2003a, 2004a, 2004b). These surveys cover seven fields: CFRS\,03, Lockman
Hole, HDF-North, SA\,13, CFRS\,14, ELAIS-N2 and SA\,22, and we discuss the
data available for each fields below. We supplement these with $IJK$
photometry of the five well-studied submm galaxies with redshifts in the
SCUBA cluster lens survey (Smail et al.\ 2002b), as given by Frayer et
al.\ (2004).

\subsection{Ground-based Optical and Near-infrared Imaging}

$I$- and $K$-band photometry (in 4$''$-diameter apertures) is available
for the sources in the ELAIS-N2 and the Lockman Hole fields from Ivison et
al.\ (2002) and we adopt the same large photometric aperture for this
work.  The choice of aperture size is driven by our desire to obtain
representative colors for the entirety of these extended (median optical
size of 2.3$''$, see \S3) and sometimes morphologically-complex galaxies
(Chapman et al.\ 2003b).  There are two penalties for this choice: first,
the apparent point-source sensitivity in the aperture is brighter than it
would otherwise be, and second in about 15\% of cases our photometry might
suffer from contamination by, probably unrelated, bright galaxies.  Where
contamination {\it may} have occured we flag these objects in our catalog.
However, we have retained these objects in our analysis as we have
confirmed that their presence does not affect any of our qualitative
conclusions.

For several of the submm sources in CFRS\,03 and CFRS\,14 fields
$I$/$K$-band photometry published in Webb et al.\ (2003) and Clements et
al.\ (2004) (but using 3$''$-diameter apertures).  We have remeasured the
photometry for these sources (and our new radio sources and submm IDs in
these fields) in 4-$''$ diameter apertures.  In addition, for the
remaining fields without published photometry (HDF, SA\,13 and SA\,22) we
have measured the 4-$''$ diameter photometry off new and available
archival imaging.  We now discuss this imaging on a field-by-field basis.

%
%
\centerline{\psfig{file=f2.ps,width=3.0in,angle=0}}

\noindent{{\sc \small Fig.~2
---}\small\small\addtolength{\baselineskip}{-3pt} A $(J-K)$--$(I-K)$
color--color plot for the SMG and OFRG galaxies in our spectroscopic
sample and the lensed SMGs with redshifts from Frayer et al.\ (2004). The
OFRGs are plotted with open symbols and all the points are coded in terms
of their redshifts. We also indicate the classification scheme proposed by
Bergstrom \& Wiklind (2004) to distinguish between extremely red galaxies
which are red by virtue of either dust (active star-forming galaxies with
redder $(J-K)$ colors) or passive, evolved stellar populations (which have
bluer $(J-K)$ colors) at $z<2.2$. Note that this only applies to the
SMG/OFRG in our lowest redshift slice.  Those SMG/OFRG at $z<2$ with
$(I-K)$ colors sufficiently red to place them in the extremely red object
(ERO) class are roughly equally divided between these two photometric
classes -- underlining the difficulty of using simple schemes to attempt
to disentangle the complex mix of obscured and unobscured activity within
the most luminous galaxies at high-$z$. }

For all of our spectroscopic sample in the ELAIS-N2 and the Lockman Hole
fields we rely on the $I$- and $K$-band imaging published by Ivison et
al.\ (2002).  For the CFRS\,03 field we have retrieved and rereduced the
$K$-band imaging used by Webb et al.\ (2003) from the CFHT archive. New
$J$/$K$-band observations of the Lockman Hole and CFRS\,03 fields were
obtained with the wide-field WIRC2 near-infrared imager (Eikenberry et
al.\ 2004) on the Palomar Hale 5-m on the nights of 2004 January 6--8.  
The seeing was 0.9--2.0$''$ and the exposure times were 3.3-ks for the
$K$-band and 6.6-ks in the $J$-band, in non-photometric conditions.
Nevertheless, the wide-field of view of WIRC2 ($8.5'\times 8.5'$) makes it
relatively simple to derive zero-points for these exposures from 2MASS
stars in these regions.  We estimate the precision of the zero-points to
be $\ls 0.04$\,mag based on the scatter between stars and the typical
depth of these observations as $J\sim 22.3$ and $K\sim 20.8$ ($3\sigma$).

%
%
\setcounter{figure}{2}
\begin{figure*}
\centerline{\psfig{file=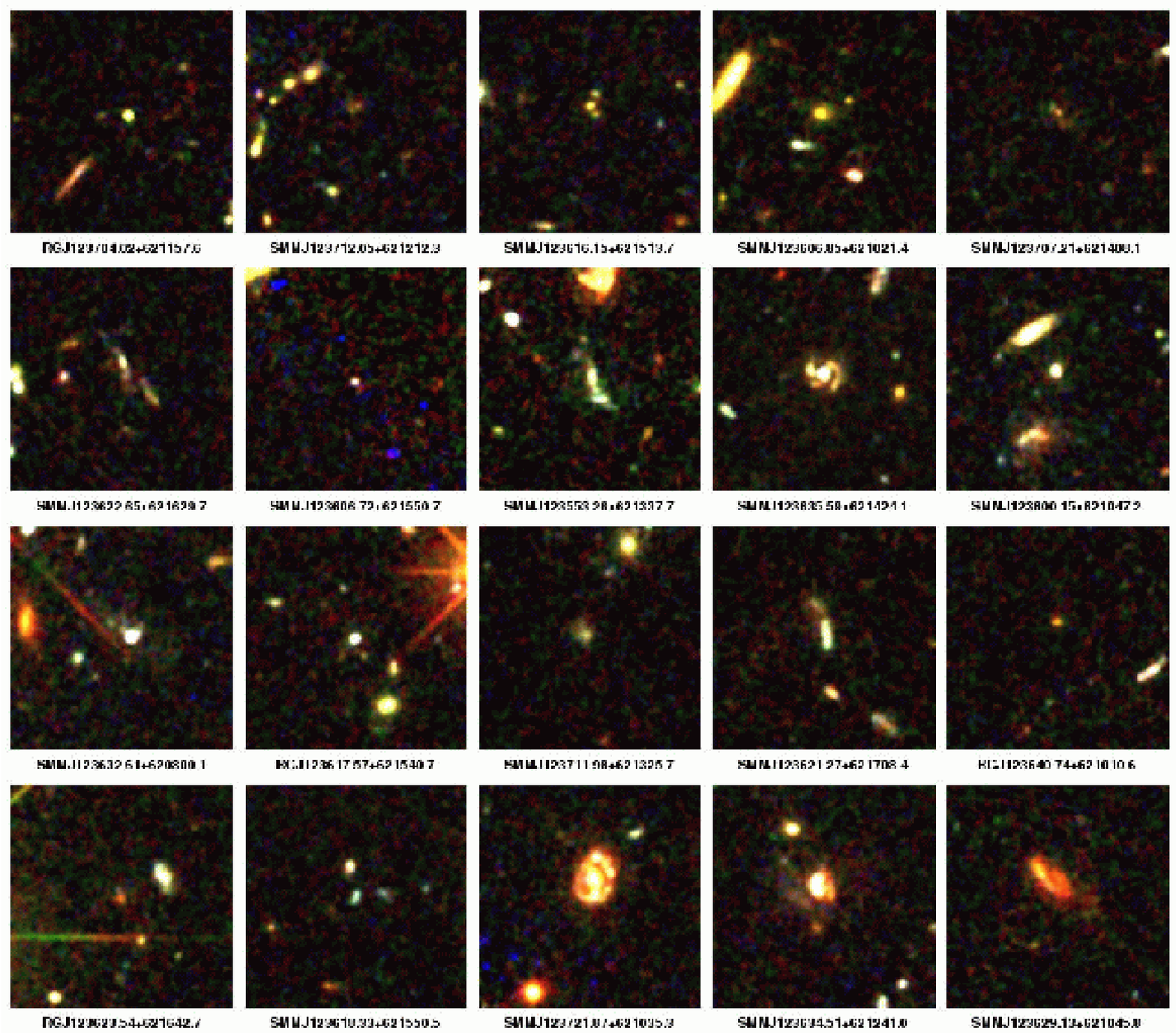,width=6.0in,angle=0}}
\caption{\small $BVI$ true-color images of 20 SMGs and OFRGs in the
GOODS/HDF-N region.  These are ordered in terms of redshift (from
$z=3.4$ at the top-left down to $z=1.0$ at the bottom-right.  The
following galaxies have $(I-K)\geq 4.0$, RG\,J123704.02,
SMM\,J123712.05, SMM\,J123606.85, SMM\,J123707.21, SMM\,J123635.59,
RG\,J123617.57, SMM\,J123621.27, RG\,J123640.74, RG\,J123623.54,
SMM\,J123618.33 and SMM\,J123629.13.  While SMM\,J123616.15,
SMM\,J123606.85, SMM\,J123606.72 and RG\,J123623.54 have bolometric
luminosities exceeding 10$^{13}$L$_\odot$.  Each panel is 7.5$''$
square, centered on the spectroscopic source and has North top and East
to the left. [Full resolution version in published manuscript]}
\end{figure*}

New $J$/$K$-band imaging of the CFRS\,14 and SA\,13 fields was also taken
with the WIRC2 imager on the nights of 2004 January 8--9.  The total
exposure times are 11.4-ks in $J$ and 3.3-ks in $K$-band for CFRS\,14 and
14.0-ks in $J$ and 10.5-ks in $K$-band for SA\,13. The conditions were
non-photometric but again the data can be reliably calibrated from 2MASS
stars serendipitous lying within the large WIRC2 field to a precision of
$\ls 0.04$\,mag. The 3-$\sigma$ limiting magnitudes of these observations
were $J\sim 22.0$ and $K\sim 20.6$.

In the HDF we make use of panoramic $K$-band imaging obtained by Bundy et
al.\ (2004) using WIRC2, as well as new $J$-band observations with the
same instrument.  The acquisition and reduction of the $K$-band data are
described by Bundy et al.\ (2004).  The observations comprise three
$8.5'\times 8.5'$ pointings completely covering the GOODS HDF-N axis with
per pointing exposure times of 15\,ks and seeing of 0.9$''$ in $K$.  We
automatically aligned and mosaiced these images using {\sc starlink
ccdpack} routines and combined them with the publically-accessible $HK$
image provided by Capak et al.\ (2003) after suitable normalisation.  The
$J$-band imaging for this field was obtained on 2004 January 6--7 with
WIRC2, under non-photometric conditions and with 1.5$''$ seeing.  Three
pointings were obtained which cover the bulk of the SMG/OFRGs in this
region, each with a total integration time of 11.6-ks in $J$. Again the
observations were mosaiced using {\sc starlink ccdpack}.  The zero points
of both the $J$ and $K$-band images are transformed onto the 2MASS
photometric system and yield 3-$\sigma$ limits in our 4$''$ photometric
aperture of $J\sim 22.5$ and $K\sim 21.0$.

In addition to the WIRC2 observation listed above, we also obtained
pointed high-resolution $J$- and $K$-band imaging of selected SMG/OFRGs
from UKIRT.  These observations cover targets in the CFRS\,03, SA\,13 and
SA\,22 fields and were obtained on the nights of 2003 February 26, August
27--31 and in queue-observing on the nights of 2003 September 8--9, 16--17
and 28 using the UFTI near-infrared imager or the UIST near-infrared
imaging spectrograph.  The typical exposure times are 4.8-ks in $J$ and
3.2-ks in $K$ with 0.5--0.6$''$ seeing and resulting 3-$\sigma$ limiting
magnitudes for point sources of $J=22.5$ and $K=20.8$.

Finally, we obtained $J$- and $K$-band imaging of the SA\,22 field using
IRIS2 on the AAT (Tinney et al.\ 2004).  These observations were kindly
undertaken by Dr.\ Mark Sullivan on the nights of 2003 September 7--9.  
These data consist of 7.2-ks in $J$ and 3.6-ks in $K$ covering a
$7.7'\times 7.7'$ field in 1.5--1.9$''$ seeing and provide 3-$\sigma$
limits of $J=21.0$ and $K=20.3$.

All the observations employ multiple short exposures (sometimes coadded
on-chip) on a dithered grid to construct a running sky-flat, which along
with suitable dark exposures is used to remove instrumental and sky
structure from the final images.  Typical per-frame exposure times were
$4\times 30$\,s in $K$ and $2\times 60$\,s in the $J$-band for the Palomar
observations, 90\,s in $K$ and 120\,s in $J$ for UKIRT and $3\times 60$\,s
in $J$ and 60\,s in $K$-band on the AAT.  The Palomar observations were
reduced in a standard manner with custom-written {\sc iraf} scripts, while
the UKIRT and AAT observations were reprocessed with the relevant {\sc
orac-dr} pipelines.

For our optical coverage we use the published $I$-band photometry from
Ivison et al.\ (2002) in the Lockman and ELAIS-N2 fields, while $I$-band
imaging for the HDF comes from the Suprimecam images published by Capak et
al.\ (2003).  The SA\,13 optical images are also taken with Suprimecam and
were retrieved from the Subaru archive and reduced with the {\sc nekosoft}
software pipeline (Yagi 1998).  We also employ $I$-band imaging of SA\,22,
CFRS\,03 and CFRS\,14 taken with the LFC camera on the Palomar Hale 5-m
and reduced with the {\sc iraf mscred} package.  Calibration in all cases
comes from observations of Landolt (1992) standard stars. Astrometry of
all of the optical (and near-infrared) imaging was tied to the USNO
catalog using interactive fits within the {\sc starlink gaia} tool.  
These astrometric solutions provide $\ls 0.5''$ rms fits to the positions
of the USNO stars in all frames.

We measure 4$''$-diameter aperture magnitudes for all the SMGs and OFRGs
from Chapman et al.\ (2003, 2004a, 2004b) lying within our $IJK$ imaging
after correcting for significant seeing differences between the passbands.  
We use this photometry to provide both {\it total} colors and magnitudes
for the galaxies and list the latter in Table~1, along with the source ID,
redshifts from Chapman et al.\ (2003, 2004a, 2004b) and any identifiers
from previous studies.  The apertures were centered on the positions of
the radio sources, but were allowed to shift by upto 1$''$ to reflect the
possible positional mismatch between the optical/near-infrared and radio
astrometry in each field.

We illustrate the distribution of the SMG/OFRG sample on the $(I-K)$--$K$
and $(J-K)$--$K$ color-magnitude planes in Fig.~1 and contrast this with
the color-magnitude distribution of the general field galaxy population
taken from a $K$-selected survey in the HDF-N (Bundy et al.\ 2004).  In
Fig.~2 we show a similar comparison of the $(J-K)$--$(I-K)$ color-color
plane for the SMG/OFRG and the field population.

%
%
\centerline{\psfig{file=f4.ps,width=3.0in,angle=0}}

\noindent{{\sc \small Fig.~4
---}\small\small\addtolength{\baselineskip}{-3pt} A comparison of the
half-light distribution for the SMG/OFRG sample at $z=1.8$--3.0 lying in
the GOODS field with the distribution for the optically-selected galaxy
population at a comparable redshift, $z\sim 2.3$, from Ferguson et al.\
(2004). The top axis shows the actual scale sizes adopting an average
angular scale of 8.3\,kpc\,arcsec$^{-1}$. The SMG/OFRG sample clearly have
much larger UV-extents than a typical galaxy selected at these redshifts
from a magnitude-limited sample. 

}

%
%
\setcounter{figure}{4}
\begin{figure*}
\centerline{\psfig{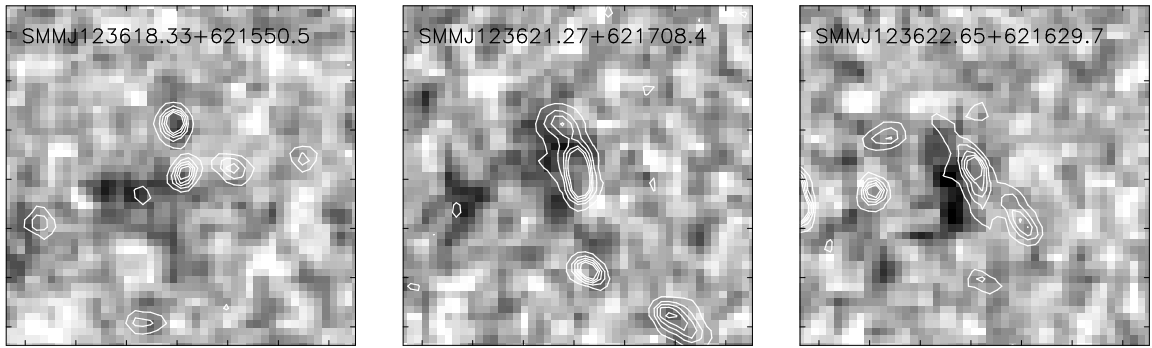}}

\caption{\small The $K$-band images, shown as a grayscale, and the
combined $BVI$ {\it HST} images (contour) for three SMGs from the GOODS
ACS region where the two wavebands show very different structures.  In
SMM\,J123618.33 the $K$-band light arises from an extremely red component
$\sim 1.5$--2$''$ ($\sim 15$\,kpc) to the East of our optically-identified
counterpart. Similarly, in SMM\,J123622.65 the near-infrared emission lies
on the edge of the optical extent of the galaxy, corresponding to a
projected offset of $\sim 5$\,kpc.  In SMM\,J123621.27, the peak in the
$K$-band light within the galaxy lies in a saddle between two
optically-bright components, while a second extremely red source is $\sim
2.5''$ ($\sim 20$\,kpc) to the East. Each panel is $7.5''\times 7.5''$ and
as the galaxies are all close to $z\sim 2$ this corresponds to $\sim
65$\,kpc. The contours are linearly spaced with an increment of 5-$\sigma$
of the sky noise.  The optical and near-infrared images have been smoothed
with 0.3$''$ and 0.5$''$ FWHM Gaussians respectively, to enhance the
contrast in the relevant structures.  The images in each passband are
locally aligned using bright, compact galaxies and have a typical error in
the alignment of less than 0.2$''$. } 
\end{figure*}

\subsection{{\it HST} Optical Observations}

In addition to the ground-based imaging discussed above, we have also
included in our analysis morphological information from higher resolution
imaging of a subset of our sample from {\it Hubble Space Telescope} ({\it
HST}).  This comprises multicolor ACS imaging of 20 galaxies from the
spectroscopic SMG and OFRG samples of Chapman et al.\ (2003a; 2004a;
2004b) which serendipitously fall within the GOODS HDF-N field.

The 5-epoch GOODS ACS F435W ($B$), F606W ($V$) and F775W ($I$) images were
downloaded from the ST-ECF mirror of the GOODS website. More details of
the acquisition and reduction of these data is given in Giavalisco et al.\
(2004).  We extract $7.5''\times 7.5''$ regions centered on the 20 SMGs
and OFRGs lying within the GOODS HDF-N imaging. We rebin these images to a
scale of 0.15$''$ pixel$^{-1}$ and then convolve them with a 0.3$''$ FWHM
Gaussian to enhance low surface brightness features before constructing
"true" color images which are shown in Fig.~3.

\medskip

In total in the final sample we have 73 SMGs, 18 OFRGs and 5 lensed SMGs,
all with redshifts.  For these we have $K$-band coverage of 89 galaxies,
with $J$-band detections or limits for 63 of the galaxies in the combined
sample.  In addition we have high-resolution {\it HST} imaging of 20
galaxies observed with ACS by GOODS.

The selection function for our combined sample is complicated, combining
as it does submm, radio, optical photometric and spectroscopic and
near-infrared photometric selection limits.  The easiest of these
selection criteria to describe is that associated with the near-infrared
imaging: which provides median 3-$\sigma$ limits of $J\sim 22.5$ and
$K\sim 20.9$.

The radio-selected SMG sample is defined by a minimum 850-$\mu$m flux of
$\sim 5$\,mJy and a 1.4-GHz flux of greater than $\sim 30\mu$Jy (Chapman
et al.\ 2004a).  Previous studies have shown that 60--70\% of a submm
sample flux limited at 5\,mJy is detectable above our radio flux limit
(e.g.\ Ivison et al.\ 2002).  The radio limit places a joint restriction
on the highest redshift and coldest dust temperature for SMGs detectable
in this survey. SMGs with characteristic dust temperatures around 40\,K
begin to fall below our radio flux limit at $z\gs 3$, with colder SMGs
missed at lower redshifts. The evolutionary models discussed by Chapman et
al.\ (2004a) suggest that the radio-selected sample is likely to be 50\%
incomplete at $z\gs 3$. As we are only considering SMGs with measured
redshifts we need to include the selection function from the spectroscopic
identifications.  Again, Chapman et al.\ (2004a) show that the bulk of the
incompleteness in the spectroscopic follow-up (25\% unidentified) likely
arises from galaxies at $z\sim 1.3$--1.8, along with weak-lined, faint
continuum sources over the full redshift range surveyed.

The OFRG sample from Chapman et al.\ (2004b) is defined by the absence of
detectable 850-$\mu$m emission (2.5-$\sigma$ limit of $<5$\,mJy)  from
optically-faint, apparently star-forming $\mu$Jy radio sources ($R>23.5$,
S$_{\rm 1.4GHz}\gs 30\mu$Jy) at $z\gs 1$. The star-formation
classification is based on their restframe UV spectral properties. At
their redshifts the radio-far-infrared correlation implies that these
galaxies will have bolometric luminosities of $\gs 10^{12}L_\odot$. As
Chapman et al.\ (2004b) discuss the lack of detectable submm emission from
these luminous galaxies is most likely explained by their hot
characteristic dust temperatures (see also Blain et al.\ 2004a).  The
OFRGs thus likely complement the SMG population, sampling the properties
of similarly far-infrared luminous galaxies with somewhat hotter dust
temperatures.

The combination of selection criteria means that we do not describe the
sample as ``complete'' in any formal sense, rather, it is representative
of the properties of a large fraction of the most far-infrared luminous
galaxies at $z\sim 1$--3.  In particular, we note that our sample is
likely to be incomplete for the optically faintest SMG/OFRGs ($I\gs 24.5$)
due to our spectroscopic survey limit. This may include the reddest
examples of the population brighter than our nominal near-infrared
magnitude limit of $K\ls 21$.  We attempt to quantify the possible biases
from our selection criteria in the next section.

Before combining the three samples we check for differences in their
photometric properties, finding median $K$-band magnitudes of
$<\!K\!>=19.70\pm 0.24$ for the SMGs, $20.23\pm 0.43$ for the OFRG and
$18.79\pm 1.28$ for the lensed SMGs (corrected to the source plane).
Similarly the median $<\!(I-K)\!>=3.71\pm 0.16$ for the SMGs, $4.14\pm
0.30$ for the OFRG and $4.00\pm 0.75$ for the lensed SMGs, while the
median $(J-K)$ colors are: $<\!(J-K)\!>=1.73\pm 0.10$, $1.91\pm 0.17$ and
$2.23\pm 0.26$.  Thus the median properties differ by less than 2-$\sigma$
for the three samples and so we feel it is justifiable to combine the
samples to study the photometric properties of luminous, dusty galaxies at
high redshift.  Nevertheless, we distinguish the OFRG and SMG samples in
the plots so that the reader can judge the behaviour of the separate
populations. The combined sample has $<\!K\!>=19.70\pm 0.23$,
$<\!(I-K)\!>=3.80\pm 0.13$ and $<\!(J-K)\!>=1.77\pm 0.06$. Removing all
those galaxies which could be affected by photometric contamination
changes these median magnitudes/colors by less than 0.3\,$\sigma$.

\section{Analysis and Results}

\subsection{Color-magnitude and color-color distributions}

Our $IJK$ photometry for 96 SMG/OFRGs (Table~1) with complete redshift
information allows us to study the restframe optical properties of a large
sample of confirmed high-redshift dusty, luminous galaxies for the first
time (c.f.\ Smail et al.\ 2002; Frayer et al.\ 2004).  We show the
color-magnitude plots for the sample in Fig.~1 and note that there is no
strong trend of apparent color with redshift, once the selection
boundaries imposed by the depth of our available imaging are taken into
account.

The primary selection criteria for inclusion in our sample is detectable
submm/radio emission ($S_{850\mu \rm m}\gs 5$\,mJy and/or $S_{\rm 1.4GHz}
\gs 30\mu$Jy). The only other major selection criterion which censors the
characteristics of the galaxies we can study are those that affect the
measurement of a redshift.  As discussed above there are two classes of
SMG/OFRG for which it is more difficult to obtain spectroscopic redshifts,
and hence maybe under-represented in our sample: both galaxies at $z\sim
1.3$--1.8, from which no strong emission or absorption features fall in
the sensitive range of the LRIS spectrograph; and galaxies without
emission lines at very faint magnitudes, $I\gs 24.5$ (see also Chapman et
al.\ 2003a, 2004a).

To determine how these selection criteria may have affected the
optical-near-infrared photometric properties of our sample we begin by
comparing the $K$-band magnitudes and $(I-K)$ colors of our spectroscopic
sample (Fig.~1), with those for the counterparts of a flux-limited submm
survey which have been precisely located from their radio emission. This
will highlight any biases which arise from our requirement for a
spectroscopic redshift. Ivison et al.\ (2002) identify counterparts to
submm sources with typical fluxes of 850$\mu$m fluxes of $\sim 8$mJy and
for those with robust radio IDs (which comprise $\gs 60$--70\% of the
whole submm population at this depth) they find that the fraction of the
population with $K\ls 21$ is 85\%. Similarly, looking at the fraction of
the population with unusually red colors, they find: $33\pm 14$\% are
extremely red objects (ERO)s with $(I-K)\geq 5.0$.  This fraction rises to
$55\pm 17$\% with $(I-K)\geq 4.0$, while only 6\% are blank (defined as
$I\geq 26$ and $K\geq 21$). In comparison, for our spectroscopic sample we
find that 90\% are detected with $K\ls 21$, $9\pm 3$\% of these have
$(I-K)\geq 5.0$ and $43\pm 7$\% have $(I-K)\geq 4.0$.  These contrast with
fractions of 3\% and 12\% respectively seen in the general $K<21$ field
population at the same median $K$ magnitude.  Thus we have a lower
fraction of the very reddest galaxies, but we appear to fairly sample
those SMG's with $(I-K)\ls 5$.  We conclude that our spectroscopic sample
of SMGs is broadly representative of the whole radio-identified submm
population.

Comparing our $(I-K)$ colors to those measured for the counterparts in a
purely submm-flux-limited sample is much more uncertain as in the absence
of either a radio or an ERO counterpart it is difficult for such surveys
to reliably identify the correct counterpart.  Webb et al.\ (2003) show
that $40\pm 15$\% of their {\it secure} sample in the CFRS\,03 and
CFRS\,14 (which includes SMGs both with and without radio counterparts)
have $(I-K)\geq 4.0$ (and $5\pm 5$\% have $(I-K)\geq 5.0$). In comparison,
Ivison et al.\ (2002) quote a fraction of $(I-K)\geq 5.0$ ERO counterparts
of 22--27\% in their full submm catalog, with between 6--43\%
blank.\footnote{Note that it has been suggested that a fraction of the
radio-blank submm sources may be spurious, Greve et al.\ (2004b).} Hence
we conclude that if our sample is biased against
optical/near-infrared--faint counterparts, then this bias is no worse than
for a radio-identified sample.

There are two published analyses of the near-infrared colors of SMGs
(Fig.~1): Frayer et al.\ (2004) and Dannerbauer et al.\ (2004). These
studies provide $JK$ photometry on small samples of lensed SMGs and
radio/mm-identified MAMBO sources. We combine these two small samples and
determine a median color of $(J-K)=2.1\pm 0.2$, with $23\pm 10$\% having
$(J-K)\geq 3.0$ and $K\ls 22.5$, and $41\pm 14$\% with $(J-K)\geq 2.3$.  
This compares with just $7\pm 3$\% (4/61) with $(J-K)\geq 3.0$ and $18\pm
5$\% (11/61) with $(J-K)\geq 2.3$ in our spectroscopic sample (and 8\% and
15\% respectively for the $K<21$ field population).  The spectroscopic
sample thus tends to be bluer than this faint, approximately flux-limited
submm sample, although the small size of the latter means these difference
are not significant.

%
%
\centerline{\psfig{file=f6.ps,width=3.0in,angle=0}}

\noindent{{\sc \small Fig.~6
---}\small\small\addtolength{\baselineskip}{-3pt} The variation of
apparent $K$-band magnitude in a fixed 4$''$ ($\sim 34$\,kpc) diameter
aperture with redshift for the SMGs and OFRGs in our sample.  We also
plot the $K$-band photometry of local ULIRGs from the 1-Jy sample of
Kim et al.\ (2002) and for more distant $>10^{12}L_\odot$ systems from
the sample of FIRST/{\it IRAS} galaxies of Stanford et al.\ (2000). In
addition we show the compilation of $K$-band photometry on powerful
radio galaxies collected by Willott et al.\ (2003).  Finally, we plot
the expected variation with redshift for a non-evolving $3 L_K^\ast$
galaxy (where $M_K^\ast = -24.3$, Kochanek et al.\ 2001), and the best fit
relation for the powerful radio galaxy sample from Willott et al.\ (2003).

}

We can also use our near-infrared imaging to determine whether we could
select the counterparts to the SMG/OFRG in our sample, {\it without} the
need for detectable radio emission.  We find that our SMG/OFRG galaxies
are either the brightest $K$-band source within an 8$''$-diameter region
around their position (a typical error box for a submm source), or if not
the brightest then typically the reddest galaxy in $(I-K)$ (Webb et al.\
2003; Borys et al.\ 2004).  For example of the 25 SMG/OFRG galaxies in the
GOODS HDF-N field, 16 are associated with the brightest $K$-band galaxy
within 8$''$, while the remaining 9 are on average $0.27\pm 0.70$ mags
redder in $(I-K)$ than any equally bright or brighter companions (with
only 3/25 or 12\% being bluer).  We caution that our spectroscopic sample
will tend to have brighter counterparts (in all bands) compared with a
purely submm-selected sample.  Hence, unfortunately, this approach does
not provide an unambigious identification of the true counterparts to
complete samples of submm sources.  Moreover, as we show below, the
counterparts across our sample already span $>6$ magnitudes in their
$K$-band brightness -- hence the identity of the true counterpart will
always remain ambigious. In part this reflects the fact that the $K$-band
emission from these galaxies is only weakly coupled to their far-infrared
emission, in contrast to their radio emission which arises from a closely
related process.  Recent mid-infrared observations from {\it Spitzer}
suggest that this may also prove to be a useful waveband for obtaining
complete identifications of counterparts to submm/mm sources (Ivison et
al.\ 2004; Egami et al.\ 2004).

The distribution of our SMG/OFRG sample on the $(I-K)$--$(J-K)$
color-color plane (Fig.~2) follows the same color-color distribution as
the general field population, with a tendency for redder colors in both
$(I-K)$ and $(J-K)$.  Such color-color plots have been discussed as a tool
to distinguish different classes of EROs (Pozzetti \& Mannucci 2000;
Bergstrom \& Wiklind 2004) and have been tested with submm and
radio-selected samples (Webb et al.\ 2003; Dannerbauer et al.\ 2004; Smail
et al.\ 2002b).  The proposed classification boundary between the passive
and dusty populations appears to split the current sample, rather than
confining it to the dusty-side of the classification region.  As our
SMG/OFRG sample are selected to be far-infrared luminous systems, it seems
unlikely that their $IJK$ colors are dominated by old, passive stellar
populations.  Rather, we view this failure as an indication that simple
two-color classification schemes may be ineffective for isolating dusty
EROs due to the complex mix of obscured and unobscured emission in these
systems. We emphasize that this failure persists when we just isolate
those SMG/OFRG EROs at $z<2$ (c.f.\ Bergstrom \& Wiklind 2004).  We also
reiterate that only a modest fraction of SMG/OFRGs have red enough colors,
$\sim 10$--30\% with $(I-K)\geq 5.0$, to be judged to be unusual (Smail et
al.\ 1999a; Webb et al.\ 2004) and hence this type of classification
scheme is only of limited use for the SMG/OFRG population.

\subsection{Morphological properties}

We show true color representations of 20 galaxies from our sample which
lie within the GOODS-N field in Fig.~3.  These are derived from the {\it
HST} ACS $BVI$ images of this field.  This sample spans $z=1.0$--3.6 and
contains 11 SMG/OFRG's with $(I-K)\geq 4.0$ but no galaxies with
$(I-K)\geq 5.0$: SMM\,J123712.05 with $(I-K)=4.95\pm 0.45$ and
RG\,J123640.74 with $(I-K)\geq 4.88$ are the reddest in the sample. Four
of these systems have estimated bolometric luminosities in excess of $\geq
10^{13}$L$_\odot$.

We can use the high-resolution optical imaging to classify the
morphologies of these reliably-identified dusty, galaxies and contrast
these with conclusions from similar studies with much less robust
counterparts (e.g.\ Smail et al.\ 1998). We find that just over half of
the sample (50--60\%) are apparently multi-component or disturbed systems,
suggestive of mergers or interactions.  However, we caution that the
presence of highly-structured dust within these galaxies might mimic such
structural pecularities (see Fig.~4 and also Smail et al.\ 1999b).  25\%
of the remaining SMG/OFRG are peculiar, including two face-on spirals, one
of which has a ring; and only 15\% are regular/normal.  Of the apparently
multi-component systems, half contain at least one component which is an
obvious disk galaxy.

The median angular scale of these galaxies is $\sim 8.4$\,kpc
arcsec$^{-1}$ and it varies by only $\pm 5$\% over their redshift range,
from 8.2\,kpc arcsec$^{-1}$ at $z=1.0$ up to 8.5\,kpc arcsec$^{-1}$ at
$z\sim 2$ and then dropping back to 7.4\,kpc arcsec$^{-1}$ at $z=3.6$.  
Hence the changing apparent size of these galaxies either reflects
variations in their true physical extents, or surface brightness dimming.  
There is clearly a trend for smaller systems at higher-redshifts, however,
the scatter at a fixed redshift is at least as large as the variation due
to this trend.  The median apparent size of the whole sample is $(2.3\pm
0.9)''$, measured at a fixed isophote (2.5\,$\sigma$ of the sky) along the
major axis of the objects in the combined $B+V+I$ images. Using their
redshifts we calculate that this corresponds to an average physical size
of $27\pm 17$\,kpc.  We also measure the half-light radii of the SMGs in
an analogous manner to that used by Ferguson et al.\ (2004), on the same
{\it HST} data, and compare these to their distribution for the general
optically-selected population at $z\sim 2.3$ (using the 15 galaxies in our
sample with $z=1.9$--3 -- giving a similar average redshift for the
subsample). We show this comparison in Fig.~4, which demonstrates that the
average SMG/OFRG in our sample at $z\sim 2.2$ is roughly twice as large as
a typical optically-selected galaxy from a similar magnitude-limited
sample at this epoch.

Our results on the morphologies and sizes of the submm population echo
those of Chapman et al.\ (2003b) -- who concluded from their sample of 13
submm-selected galaxies with STIS or WFPC2 imaging, that 83\% have
irregular morphologies and just 17\% had regular structure (this is almost
identical to the $85\pm 10$\% and $15\pm 2$\% for our sample). Chapman et
al.\ (2003b) also compared the UV sizes of their SMGs to samples of LBGs
and magnitude-matched optically-selected galaxies and suggested that the
SMGs were larger than either of these comparison samples.  The
availability of redshifts for our sample has allowed us to demonstrate
that these dusty, luminous systems are on average twice as large as the
typical galaxy at their epoch.

We have also compared the observed $K$-band and optical morphologies of
all the galaxies in the GOODS HDF-N region to search for more obscured
components within these systems.  We find that $\sim 20$\% (3/18) of the
sample display near-infrared morphologies which strongly deviate from
those seen in the optical.  We show these three galaxies in Fig.~5 --
where we see that there are extremely red components within some of these
systems, with $(I-K)\geq 5$, on scales from 5--20\,kpc.  We also find
evidence for dust extinction within some of the systems, which may help to
explain the apparently disturbed, multi-component structures of these
galaxies.  The near-infrared morphologies of some of our SMG/OFRGs show
strong similarities to those of high-redshift luminous radio galaxies
(Pentericci et al.\ 2001), hinting at a similar mode of triggering for the
activity in these two populations. The presence of very red components on
$\sim 2''$ scales within some SMG/OFRGs underlines the need for
large-aperture photometry to derive reliable colors for the whole system
(c.f.\ Dannerbauer et al.\ 2004).  We will discuss the internal structure
of the SMGs on $<1$\,kpc scales in a future paper using a sample of {\it
HST} NICMOS and ACS imaging of galaxies from our spectroscopic SMG/OFRG
sample.

\subsection{Magnitude--redshift and color--redshift distributions}

We now begin to investigate the variation of galaxy properties with
redshift within our sample.  We show in Fig.~6 the $K$-band Hubble diagram
for our sample of dusty, luminous galaxies and compare this to the
behaviour of local and more distant ULIRGs, as well as samples of
powerful, high-redshift radio galaxies.

We note that we see no trend of $K$-band magnitude with $S_{850\mu\rm m}$
within the SMG/OFRG sample. This is at odds with Smail et al.\ (2002b),
who suggested galaxies which were fainter submm sources might also be
fainter in $K$-band (but see Webb et al.\ 2003).  However, we caution that
our current sample is restricted in the brightness of the counterparts by
the need to measure a spectroscopic redshift and hence is not ideal for
performing this test.  In particular, the 5-mJy submm flux limit for our
SMG sample means that we have only marginal overlap with the flux regime
studied by Smail et al.\ (2002).

Turning back to the $K$--$z$ relation in Fig.~6, and discarding the two
bright $K$-band counterparts which are obvious broad-line AGN, we see that
the bright wing of the SMG/OFRG distribution matches that traced by
high-redshift luminous radio galaxies, roughly corresponding to $\sim
5L^\ast$.  However, in contrast to the tight radio galaxy distribution,
the SMG/OFRG sample spans 4--5 magnitudes range in $K$-band brightness at
a fixed redshift, and roughly tracks the predicted locus of an unevolving
$3L^\ast$ elliptical galaxy.  The large variation in the $K$-band
magnitudes of SMG/OFRG is perhaps unsurprising given that many of these
galaxies appear to be multi-component systems, with strong internal
extinction and young starburst ages (Smail et al.\ 2003a).

%
%
\centerline{\psfig{file=f7.ps,width=3.0in,angle=0}}

\noindent{{\sc \small Fig.~7
---}\small\addtolength{\baselineskip}{-3pt} The distribution of
SMG/OFRG on the $(I-K)$--$z$ plane.  We compare this to the predicted
$(I-K)$ colors as a function of redshift of three local ULIRGs derived
from their observed restframe UV SEDs by Trentham et al.\ (1999). We
see that there is a wide range in apparent colors of SMG/OFRG at a
fixed redshift, in line with that expected from the restframe UV colors
of local ULIRGs.  We believe the diversity of colors probably reflects
a wide range of internal obscuration and mix of components within these
galaxies -- the presence of multiple components and highly-structured
dust is also indicated by the UV and near-infrared morphologies of
these galaxies (Figs.~3 \& 5).  We scale the size of each point with
the reddening-corrected absolute $V$-band magnitude of the galaxy,
assuming a constant star formation rate model.  For completeness we note
that there is no significant correlation of median colors with
850-$\mu$m flux and no correlation of 850-$\mu$m flux with redshift
(Chapman et al.\ 2004a). 

} 

We fit a simple polynomial relation to the variation of $K$-band apparent
magnitude with redshift for the SMG/OFRG sample (having removed the two
obvious luminous $K\sim 16$ AGN at $z>2$) and derive a best-fit relation
of $K=18.51+2.90\times \log_{10}(z)+2.396\times (\log_{10}(z))^2$ with a
scatter of $\sigma_K =0.97$\,mags around this.\footnote{Removing those
objects whose photometry may be contaminated by near neighbors does not
significantly reduce the scatter in this fit. Similarly, removing
SMM\,J030238.62+001106.3, the outlier at $z=0.28$/$K\sim 20.5$, only
reduces the scatter to $\sigma_K = 0.89$\,mags. We believe that the
redshift for this source probably refers to an unrelated foreground dwarf
galaxy, with the true SMG counterpart being even fainter and at $z>1$.}
This scatter is 60\% higher than that seen in the luminous radio galaxy
population at similar redshifts (Willott et al.\ 2003).  The SMG/OFRG
sample also shows an offset of $\Delta K = 0.95\pm 0.19$ from the
radio-galaxy $K$--$z$ relation of Willott et al.\ (2003) -- indicating
that the SMG/OFRG are intrinsically less luminous in the restframe
$V$-band than powerful radio galaxies at the same redshift.  Applying the
reddening estimates from the photometric modeling (\S3.5) reduces the
scatter in the SMG/OFRG $K$--$z$ relation by $\ls $\,10--20\%: although
the fainter galaxies tend to be bluer in $(I-K)$ at a fixed-$z$ (Fig.~7)
and so have lower estimated reddenings.  Looking at the scatter around
this fit for the 11 galaxies with bolometric luminosities of
$>10^{13}L_\odot$ and $K$-band coverage, we find only a marginally smaller
dispersion $\sigma_K = 0.90$\,mags.  If we fit the polynomial to just
these 11 galaxies the dispersion of the fit drops marginally to $\sigma_K
= 0.74$\,mags, but a Monte Carlo simulation indicates that this reduction
in scatter is not statistically significant, with the fits for 17\% of
random samples of 11 SMG/OFRGs from our catalog showing dispersions below
0.74\,mags (c.f.\ Serjeant et al.\ 2003).  A definitive conclusion about
the tightness of the $K$--$z$ relation for the most luminous galaxies will
have to await a larger and more homogeneous sample than either of those
used so far.

We can also invert the $K$-band Hubble diagram and ask how precisely we
could estimate the redshifts for an SMG/OFRGs given an apparent $K$-band
magnitude.  We determine a simple linear relation with $z=-4.28+0.33\times
K$ with $\Delta z/z=0.41$ (again after discarding the two bright AGN).  
Hence near-infrared photometry provides only a very crude measure of the
redshift for SMG/OFRG.

Finally, we look at the variation in the optical-near-infrared colors of
the sample with redshift.  As Fig.~7 shows, we find no discernable trends
in the median $<\! (I-K)\! >$ (or $<\! (J-K)\!>$) with $z$, with the
gradients of fits to the median values of three bins with equal numbers of
galaxies consistent with zero. The trends of color with redshift are
roughly bracketed by the predicted tracks for three local ULIRGs from
Trentham et al.\ (1999), once allowance is made for our likely
incompleteness for the reddest galaxies (\S3.1).  However, there is a
substantial fraction of the population which are even bluer in $(I-K)$
than the bluest of the three ULIRGs observed by Trentham et al.\ (1999),
although this is hardly surprising given their small sample.  We attribute
the wide variation in restframe UV--optical colors of SMG/OFRGs to
differing levels of obscuration and mixes of unobscured and obscured
components within the galaxies.  In part this variety may come about as a
result of different ages for the activity within the systems, with systems
becoming less obscured as they age, or by differences in the porosity of
the obscuring material brought about through the actions of winds and
outflows.  Less-obscured channels would provide an essential conduit for
the escape of the Ly$\alpha$ emission frequently identified in the spectra
of this purportedly dusty population (Chapman et al.\ 2003, 2004a, 2004b).

\subsection{Comparison to other high-$z$ populations}

Recently a number of studies have been published dealing with the $(J-K)$
colors of faint field galaxies. A central claim of this work is that
photometric selection in this color, specifically $(J-K)\geq 2.3$, is a
reliable a route to identify galaxies at $z\gs 2$ (Franx et al.\ 2003;
Daddi et al.\ 2003).  The $\sim 20$\% (11/61) of the SMG/OFRG with
$(J-K)\geq 2.3$ in our sample have median properties of $K=19.29\pm 0.22$
and $z=2.25\pm 0.15$, with one example at $z=1.01$ and the remaining ten
at $z=1.99$--2.62.  This confirms that, at least for these dusty, active
galaxies, a cut at $(J-K)\geq 2.3$ can cleanly select high-$z$
counterparts to SMG/OFRGs.  For comparison a $(J-K)\geq 2.0$ selection
leads to 30\% of the sample at $z\ls 1.0$--1.5.

To estimate how large a fraction of the photometrically-selected
$(J-K)\geq 2.3$ population consists of SMG/OFRG we first estimate the
median submm flux for this subsample, finding $S_{850\mu\rm m} = 4.7\pm
1.0$\,mJy.  Submm sources with this flux density have a corresponding
surface density of $\sim 0.5$ arcmin$^{-2}$ and hence the spectroscopic
$(J-K)\geq 2.3$ fraction in our sample has an effective surface density of
$\geq 0.1$ arcmin$^{-2}$, representing only about 6\% of the overall
$(J-K)\geq 2.3$ population with $K<20.5$. The proportion of submm-bright
$(J-K)\geq 2.3$ galaxies may correspond to at most 10\% of this
photometric class, suggesting that the overlap between the
photometrically-selected sample and SMG/OFRGs is at best modest (see also
Dannerbauer et al.\ 2004).  A similar analysis of UV-selected
high-redshift samples by Chapman et al.\ (2000), confirms that extreme
near-infrared colors do not appear to isolate the far-infrared luminous
fraction of the high-redshift galaxy population.  However, given their
apparently large stellar masses (and strong clustering, Daddi et al.\
2003) it seems likely that the passive and recently star-forming galaxies
with very red near-infrared colors at $z\sim 2$ could be immediate
descendents of the short-lived $z>2.5$ SMG/OFRG population.

%
%
\centerline{\psfig{file=f8.ps,width=3.0in,angle=0}}

\noindent{{\sc \small Fig.~8 ---} \small\addtolength{\baselineskip}{-3pt}
The restframe SEDs of those SMG/OFRGs which are detected in both the $J$-
and $K$-bands (73\% of those with $J$-band observations).  These are
arbitrarily normalised based on their monochromatic flux at 0.5\,$\mu$m.  
We show the SEDs of the individual galaxies and also the median of these
as a function of wavelength, along with the 1-$\sigma$ scatter on this
estimate.  The median SED shows a very red continuum shape with a
discontinuity in the continuum slope around $\sim 0.4\mu$m.  We compare
this to two simple model SEDs: for an evolved stellar population, a
luminous local elliptical galaxy; and a dusty star-burst, based on a
100-Myr old constant star formation model with $A_V=4$ from Leitherer et
al.\ (1999).  The dusty, young star-forming galaxy appears to provide a
better match to the observed SED and suggests that the discontinuity in
the continuum slope is associated with the Balmer break (highlighted by
the shaded region).

}

We can also compare the photometric properties of our SMG/OFRG sample with
samples of UV-selected star-forming galaxies identified at similar
redshifts: the Lyman-break galaxies (LBGs). For a sample of 18 SMG/OFRGs
cut in redshift to match the redshift distribution and median redshift of
LBGs in Shapley et al.\ (2001), $z=2.6$--3.4, we find $<\!K\!>=20.49\pm
0.20$, $<\!(I-K)\!>=3.54\pm 0.38$ and $<\!(J-K)\!>=1.73\pm 0.11$.  By
comparison, Shapley et al.\ (2001) LBG's have $<\!K\!>=21.35\pm 0.14$
($N=42$), $<\!(I-K)\!>=2.40\pm 0.10$ after transforming from $(R_{AB}-K)$
using $(R_{AB}-I)\sim0.4$ appropriate for galaxies with their observed
$(R_{AB}-K)$ colors at $z>2.6$, and $<\!(J-K)\!>=1.63\pm 0.10$.  Hence the
SMG/OFRGs lying within the redshift range of the classical LBG selection
are typically twice as bright in the $K$-band (restframe $V$-band) and
slightly redder in their restframe UV and optical colors.  However, we
note that our radio-selected sample would be incomplete for the faintest
SMGs if there is a strong correlation between $K$-band and radio
luminosities, leading to the detected SMGs being brighter, but bluer, than
would be seen in a pure submm-selected sample.

The classical LBG samples lie at somewhat higher redshifts than the
majority of our SMG/OFRG sample (this may in part be responsible for the
poor overlap between these two populations, Chapman et al.\ 2000). Recent
work by Steidel et al.\ (2004) has extended their photometric selection
techniques to lower redshifts, $z=1.4$--2.6, giving a better match for
comparison to the bulk of the SMG/OFRG population.  These ``BX/BM''
samples (mean $z =2.23\pm 0.31$) are brighter and redder than the LBG
population: $<\!K\!>=20.49\pm 0.06$ (for $K<21$) $<\!(I-K)\!>=3.22\pm
0.05$ (adopting $(R_{AB}-I)\sim0.35$).  The equivalent measurements for
the 57 SMGs/OFRGs at $z=1.4$--2.6 (mean $z=2.15\pm 0.30$) in our sample
are $<\!K\!>=19.77\pm 0.29$ and $<\!(I-K)\!>=3.99\pm 0.16$.  Again we find
that the SMG/OFRG are both brighter and redder than the UV-selected
population at $z\sim 2$.  The difference in $K$-band magnitude amounts to
roughly a factor of two difference between the typical restframe $R$-band
luminosities of these two populations. Without more information it is
impossible to determine whether this results from the larger stellar
masses of the SMG/OFRG sample, or to a strong contribution to their
restframe $R$-band fluxes from the H$\alpha$ line which will fall in the
observed $K$-band at their typical redshifts. Swinbank et al.\ (2004)
discuss near-infrared spectroscopy of a sample of SMG/OFRG and conclude
that the H$\alpha$ line typically contributes only about 10\% of the
broad-band flux and so we suggest that the difference in mean
$K$-magnitude probably reflects the brighter continuum luminosity of the
SMG/OFRG galaxies.  Similarly, it is unlikely that emission-line
contributions strongly skew the observed colors of the SMG/OFRG,
suggesting that the SMG/OFRG have redder continua than the average BX/BM
galaxy -- although without more detailed modeling it is impossible to tell
whether this is due to either more evolved stellar populations or dust
reddening.  Mid-infrared photometry of spectroscopically-identified
SMG/OFRG from {\it Spitzer} should provide a powerful tool to distinguish
between these alternative explanations.

%
%
\centerline{\psfig{file=f9.ps,width=3.0in,angle=0}}

\noindent{{\sc \small Fig.~9 ---} \small\addtolength{\baselineskip}{-3pt}
The ratio of restframe $V$-band luminosity to far-infrared luminosity as a
function of far-infrared luminosity for the SMG/OFRG sample.  We also show
the distribution of the local ULIRGs in the 1-Jy ULIRG sample of Kim et
al.\ (2002) and the luminous infrared galaxy sample of Veilleux et al.\
(1995), converted from $M_R$ and $M_B$ respectively to $M_V$ using median
colors of $(V-R)=0.9$ and $(B-V)=1.0$ corresponding to SEDs with their
observed $(R-K)$ colors at $z\sim 0.15$. 

}

\subsection{Simple photometric modeling}

To try to constrain the extinction and ages of the activity in the
SMG/OFRG population we have sought to exploit the combination of
broad-band photometry and redshifts available for our sample.  We start by
simply plotting the restframe spectral energy distributions (SEDs)  of the
46 galaxies with $J$- and $K$-band detections in Fig.~8.  If the galaxies
have similar restframe SEDs then their spread in redshift enables us to
recover information about the SED shape on a finer scales than provided by
the broadband colors. We have therefore also calculated the median
restframe SED of this population and show this in Fig.~8.  The median SED
shows a red continuum from $\sim 0.4$--0.7\,$\mu$m, with an apparent break
to a steeper continuum slope below $\sim 0.4\mu$m.  We compare this to the
SED of an evolved stellar population and a young, highly-reddened star
forming galaxy from {\sc starburst99} (Leitherer et al.\ 1999).  The
continuum slope in the red and the position of the break in slope are both
more consistent with the dusty, starburst model -- with the discontinuity
in the continuum slope arising from the Balmer break.  However, we caution
that the $J$-band lies at $\sim 0.4\mu$m at the median redshift of our
sample and hence the strength of the apparent discontinuity may change if
we included those galaxies with only limits on their $J$-band fluxes.

To provide a more quantitative analysis of the colors of SMG/OFRGs we have
fitted the observed $IJK$ photometry of the galaxies at their observed
redshifts using {\sc hyper-z} (Bolzonella et al.\ 2000) to derive
reddening estimates for the stellar populations in these systems, assuming
that they are dominated by young starbursts (consistent with their intense
far-infrared emission and the median SED derived above).  For the
underlying star formation histories we adopt either a single burst or
constant star formation rate (SFR) model and a Calzetti reddening law with
$A_V\leq 5.0$.  We use a likelihood ratio test to select the best-fitting
model at the known redshift, indicating that 66\% are better fit by the
instantaneous burst models. We derive a mean age of the stellar
populations in the $z>1$ SMG/OFRG sample of $450\pm 80$\,Myr and a mean
reddening of $A_V=1.70\pm 0.14$. This reddening is consistent with the
range measured for the nuclear regions of local ULIRGs (Scoville et al.\
2000). We also find, as expected, that the derived reddening correlates
with $(I-K)$ at a fixed redshift and that the best fit ages are older at
lower redshifts.

While the photometry for the majority of SMG/OFRGs are well fit by these
simple models, with 85\% having probabilities for the fits of $\geq 95$\%,
we caution that the constraints on individual galaxies are very weak and
so in our discussion we will focus on the ensemble properties of the
sample (although we stress that the availability of precise redshifts
removes much of the ambiguity in this analysis).  We also note that the
worst-fitting sources include several bright AGN and that the galaxies
with poorly-fit SEDs are uniformly distributed across the fields in our
survey. To investigate the systematic errors in these estimates we fit
either just a single burst model {\it or} a constant SFR model and derive
mean ages and reddenings of $310\pm 90$\,Myrs and $A_V=1.70\pm 0.14$ or
$530\pm 80$\,Myrs and $A_V=2.44\pm 0.13$ respectively.  The relatively old
ages for the single burst model are inconsistent with the detection of
strong far-infrared emission from star formation in these galaxies, and so
we interpret them rather as an indication of past star formation activity
within these galaxies (on 100's Myr timescales).  For this reason we
prefer to use the reddenings and luminosities derived from the constant
star formation model fits in the analysis that follows.

To attempt to compare the results of this modelling between populations,
we have also applied this same fitting technique to the $RJK$ photometry
of the $z\sim 3$ LBGs in Shapley et al.\ (2001). Here we derive a mean age
of $250\pm80$\,Myr and a mean extinction of $A_V=0.92\pm 0.14$.  
Restricting ourselves to only a constant SFR model and requiring that the
derived ages are in excess of 10\,Myrs, we measure a mean age of $330\pm
100$\,Myrs and reddening of $E(B-V)=0.32\pm0.05$, roughly consistent with
the $590\pm 60$\,Myrs and $E(B-V)=0.17\pm 0.01$ determined from a more
careful analysis of their $GRJK$ colors by Shapley et al.\ (2001).  As we
noted earlier, Steidel et al.\ (2004) have already shown that $z\sim 3$
LBGs are bluer/fainter in their restframe than the $z\sim 2$ BX/BM
population -- so it would be informative to repeat this comparison when
near-infrared photometry becomes available for the latter.

Given the uncertainties in this analysis and our limited dataset we
conclude that the restframe UV/optical emission from the SMG/OFRG
population is dominated by the light from a highly reddened and very young
stellar population, with an age of just a few 100's Myrs and typical
continuum extinction corresponding to a factor of $\gs 50\times$
obscuration at 1500\AA.  We estimate the reddening for the SMG/OFRG to be
at least twice that of the LBGs we have analysed, and perhaps more as a
result of the degeneracy in the fits between older ages and increased
reddening.  This indicates that the redder optical/near-infrared colors we
measured in \S3.4 for the SMG/OFRG sample result from stronger extinction
compared to the LBGs and BX/BM populations.  Correcting for this stronger
extinction would suggest that the SMG/OFRG's have dust-corrected,
restframe optical luminosities at least five times brighter than typical
BX/BM or LBG systems.

%
%
\centerline{\psfig{file=f10.ps,width=3.0in,angle=0}}

\noindent{{\sc \small Fig.~10 ---}\small \addtolength{\baselineskip}{-3pt}
The absolute restframe $V$-band magnitudes of SMGs and OFRGs with
redshifts lying in $z=1$--2.5 (beyond which our $K$-band imaging becomes
increasingly incomplete for counterparts with $M_V\ls -20$).  We overplot
fits to the $M_V\leq -20$ data points (shown as filled symbols) for both
Schecter and Gaussian function fits.  For the Schechter function fit we
derive $\alpha=-1.03\pm 0.27$ and $M_V^\ast=-23.1\pm 0.6$.  From a
Gaussian fit we estimate $<M_V>=-21.1\pm 0.25$ and $\sigma_V=0.75\pm 0.1$.
We also show the restframe $V$-band luminosity function derived for $z\sim
3$ LBGs by Shapley et al.\ (2001) and that of local ULIRGs from the 1-Jy
sample.  These demonstrate that the typical SMG/OFRG has a similar
restframe optical luminosity to ULIRGs at $z\sim 0$ (although it is
several times brighter in the far-infrared) and has a characteristic
luminosity which is $4\times$ brighter than the somewhat higher-redshift
LBG population (correcting for dust extinction will exacerbate this
difference).  

}

We can also test the consistency of the estimated ages and luminosities by
determining whether it is possible to form the observed stellar population
given the estimated star formation rate for the galaxies within the
proposed timescale.  We take the median $V$-band luminosity from the {\sc
hyper-z} SED fits and correct this on a case-by-case basis with the
estimated reddening for each galaxy, giving us a median dereddened
$V$-band luminosity of L$_V\sim 1.8\times 10^{11} $L$_\odot$.  For a
system with a constant star formation rate and an age of a few 100's Myrs
we require an SFR of $\ls $1000\,M$_\odot$\,yr$^{-1}$ (for stars more
massive than 1\,M$_\odot$) to build up this luminosity. The median
far-infrared luminosities of the $z>1$ SMG/OFRG sample is $5\times
10^{12}L_\odot$ corresponding to an SFR of $\sim 2\times
10^3$\,M$_\odot$\,yr$^{-1}$ for stars more massive than $>1$\,M$_\odot$.  
This suggests that the star formation activity in the average SMG/OFRG is
more than sufficient if it was maintained at the current level to produce
the observed optical luminosities within the last 100\,Myrs.  However, we
feel it is more likely that the activity is episodic, probably comprising
short $\ll 100$\,Myr bursts spread over several 100's Myrs. Such behaviour
would be consistent with that expected from merger-driven starbursts
(Mihos \& Hernquist 1996), as well as the apparent masses of black holes
seen in some SMGs (Smail et al.\ 2003a, 2003b).

We can also use the estimated mass-to-light ratios of young starbursts
derived using {\sc starburst99} (Leitherer et al.\ 1999) to estimate a
minimum stellar mass for these systems: this gives an average mass to
light ratio of $M/L_V\sim 0.15$\,M$_\odot/L_\odot$ for a system with
either a constant SFR over a few 100\,Myrs or a $\ls 300$\,Myr starburst.  
Applying this to our reddening-corrected median $V$-band luminosity, we
estimate that the typical SMG/OFRG contains $M_{\rm stars}\sim 3\times
10^{10}$M$_\odot$ of young stars, in addition to any underlying older,
more evolved population. Furthermore, millimeter-wave CO mapping of a
handful of SMGs (Frayer et al.\ 1998, 1999; Neri et al.\ 2003) indicates
that the median gas mass in submm-luminous galaxies is of order $M_{\rm
gas}\sim 2\times 10^{10}$M$_\odot$.  This suggests that many of our
SMG/OFRG retain enough gas to continue forming stars at the current rate
for a similar length of time to their current ages.  Combining the stellar
and gas masses gives a typical baryonic mass for an SMG in our sample of
$M\sim 5\times 10^{10}$M$_\odot$. This is comparable to the stellar mass
of an L$^\ast$ galaxy at the present-day (Cole et al.\ 2001), indicating
that over half our sample are likely to leave $\gs L^\ast$ descendents at
the present-day (Genzel et al.\ 2003).

\subsection{Comparison to $z\ll 1$ ULIRGs}

There is a class of obscured, active galaxies at relatively low redshifts
which have almost comparable far-infrared luminosities to those estimated
for the SMG/OFRG population: the ULIRGs.  If we can demonstrate that
SMG/OFRGs share the same properties as ULIRGs, then we can use the much
more detailed observational information available on the latter to infer
more about the processes which may operate in the more distant population.

To compare the properties of the SMG/OFRG sample with similarly luminous
low-redshift galaxies we begin by using the restframe $V$-band
luminosities from the photometric modeling described in the previous
section.  We stress that at the typical redshift of our SMG/OFRG sample
the restframe $V$-band is straddled by our observed $JK$ photometry making
this estimate of their optical luminosities relatively straightforward. We
plot the ratio of the $V$-band and far-infrared luminosities for our
sample as a function of their far-infrared luminosities in Fig.~9.  We
compare this distribution to the equivalent measurements for the samples
of ULIRGs from Kim et al.\ (2002) and luminous infrared galaxies from
Veilleux et al.\ (1995).  For comparison, the median far-infrared
luminosities are $(1.9\pm 0.2) \times 10^{12} $\,L$_\odot$ for the 1-Jy
sample, converted to our cosmology, $(3.7\pm 0.4) \times 10^{12}
$\,L$_\odot$ for $z= 1$--2.5 SMG/OFRG and $(5.0\pm 0.6) \times 10^{12}
$\,L$_\odot$ for the full SMG/OFRG sample. Together these samples exhibit
a broad trend of lower L$_V$/L$_{\rm FIR}$ at higher luminosities, which
has been interpreted as a signature of an increasing fraction of highly
obscured star formation in more active systems (e.g.\ Serjeant et al.\
2003).

As expected, the lower luminosity (usually lower redshift) systems within
the SMG/OFRG sample are distributed within the broad correlation seen in
the local far-infrared-selected samples, indicating that these galaxies
are likely to be entirely analogous to similarly luminous, low redshift
systems selected from {\it IRAS} surveys.

The more typical, higher luminosity and higher redshift, SMG/OFRG extend
the trends seen in the lower-luminosity samples to lower L$_V$/L$_{\rm
FIR}$ in more luminous galaxies, although with two orders of magnitude
scatter at a fixed luminosity.  To quantify the trend we see, we calculate
the median L$_V/$L$_{\rm FIR}$ of the $z=1$--2.5 SMG/OFRG (L$_V/$L$_{\rm
FIR} = 0.005\pm 0.001$) and for the {\it IRAS} 1-Jy sample of Kim et al.\
(2002), L$_V/$L$_{\rm FIR} = 0.023\pm 0.003$, showing that the SMG/OFRG
are roughly four times more obscured than local ULIRGs (Fig.~9).

We have also used the restframe $V$-band absolute magnitudes for the
SMG/OFRG to construct a luminosity function for an approximately
volume-limited sample of SMG/OFRG brighter than $M_V= -20$ at $z=1$--2.5.
This is shown in Fig.~10 and exhibits a steep rise at $M_V\gs -24$, with a
decline at $M_V\gs -20$ which we attribute to incompleteness.  We compare
it to the equivalent distribution for the 1-Jy sample of local ULIRGs (Kim
et al.\ 2002), which has a median absolute $V$-band magnitude of
$M_V=-21.15\pm 0.09$, converted from $M_R$ using a typical color of
$(V-R)=0.9$ suitable for a galaxy SED at $z=0.15$ with $(R-K)=3.25$. The
galaxies in the $z=1$--2.5 SMG/OFRG sample have an almost identical median
absolute magnitude: $M_V=-21.05\pm 0.27$. Thus the absolute $V$-band
luminosities of the local ULIRGs and the distant SMG/OFRG are very similar
(Fig.~10), indicating that the factor of four difference in their
L$_V$/L$_{\rm FIR}$ ratios (Fig.~9) is most likely due to the higher
far-infrared luminosities of the distant galaxies.  This is consistent
with a constant star formation efficiency for low- and high-redshift
starbursts (c.f.\ Baugh et al.\ 2004), given the factor of 3 times higher
gas fractions and larger gas masses seen in the distant SMG/OFRGs (Frayer
et al.\ 1999;  Neri et al.\ 2003).

\subsection{Evolution of SMGs}

As the final step in our analysis we wish to understand the evolution of
SMG/OFRG to the present day.  To do this we take one extreme scenario --
and assume that the star formation activity we see is the last major event
in these galaxies and that their subsequent evolution can be approximated
by passive evolution.  We can then employ simple stellar evolution models
to predict the likely luminosity function of this population at lower
redshifts and compare that to possible descendent populations.  As we
discussed in \S3.5, the large gas reservoirs detected in SMGs through
their CO emission suggests that these galaxies retain enough gas to
modestly increase their stellar masses.  However, for the purposes of this
simple discussion we will ignore any additional sources of new stars.

To quantify our discussion, we use the characteristics of the $z>1$
SMG/OFRG sample from our earlier analysis. We correct the restframe
$V$-band luminosities of the galaxies using their individual estimates of
$A_V$ from the constant SFR fits and derive a reddening-corrected absolute
magnitude of $M_V=-23.35\pm 0.15$ for a sample with a median redshift of
$z\sim 2.2$.  The luminosity-weighted ages we derived from our {\sc
hyper-z} fits were a few 100's Myr, consistent with the expected ages of
the starbursts in these massive galaxies from dynamical arguments.  We now
turn to the {\sc pegase} spectral modeling package (Fioc \&
Rocca-Volmerange 1997) to estimate the restframe $V$-band fading of a
system with a constant star formation for $\sim 100$--500\,Myrs
(bracketing the likely range in ages), which is terminated at that point
and is observed 5.3\,Gyrs later (corresponding to the time between $z\sim
2.2$ and $z=0.55$). We find that the stellar populations are expected to
fade by between $\Delta V = -3.5$ and $-4.5$ magnitudes and so we take
$\Delta V \sim -4$ as representative. Thus, the typical SMG/OFRG will have
$M_V\sim -20$ at $z=0.55$ assuming there are no subsequent star formation
events.\footnote{We have chosen to undertake this comparison with $z=0.55$
clusters as the passive galaxies populating the faint end of the
luminosity function in more local clusters may be built up through a
different process, involving passive fading or more active
threshing/harassment of mid-/late-type disk galaxies accreted by the
clusters since $z\sim 0.5$--1 (Dressler et al.\ 1997; de Lucia et al.\
2004; Kodama et al.\ 2004).  Hence this comparison is most easily achieved
with a distant cluster where the luminous elliptical population can be
more easily isolated.}

The proposed evolution of the luminosity function for SMG/OFRGs is
illustrated in Fig.~11.  This shows the observed restframe $V$-band
absolute magnitude distribution of the $z>1$ SMG/OFRG sample, the same
distribution corrected on an object-by-object basis for the estimated
reddening, and then finally this distribution faded by 4 magnitudes to
represent the passive evolution from $z\sim 2.2$ to $z=0.55$. The final
distribution is compared to that of morphologically-classified spheroidal
galaxies (almost all of which are ellipticals) in three $z\sim 0.55$
clusters from the study of Ellis et al.\ (1997) (with arbitrary
normalisation). This comparison suggests that in this simple evolutionary
picture the SMG/OFRG population can adequately match the bright end of the
luminosity function of spheroidal galaxies seen in cluster environments.  
This is clearly highly speculative, but taken in conjunction with the
large stellar masses we derive for these galaxies, their large dynamical
masses (Swinbank et al.\ 2004; Neri et al.\ 2003; Greve et al.\ 2004a) and
their strong clustering (Blain et al.\ 2004a), it provides circumstantial
evidence for the association of the SMG/OFRG with the early formation of
the most luminous ($>L^\ast$) spheroidal galaxies.

%
%
\centerline{\psfig{file=f11.ps,width=3.0in,angle=0}}

\noindent{{\sc \small Fig.~11 ---}
\small\addtolength{\baselineskip}{-3pt}This plot shows the observed
restframe $V$-band luminosity function for the $z=1$--2.5 SMG/OFRG sample
along with a best-fit Gaussian function, the same distribution corrected
for reddening using the individual extinction corrections estimated from
the {\sc hyper-z} fits to their $IJK$ colors, also overlayed with its
best-fit Gaussian function and finally this same functional fit, after
fading by $\Delta V \sim 4$ magnitudes corresponding to passive evolution
for 5.4\,Gyrs starting with a galaxy with a constant star formation
history and an age of 100--500\,Myrs at $z=2.2$.  This predicted
luminosity distribution is compared to the observed luminosity function of
morphologically-classified E/S0 galaxies in three $z\sim 0.55$ clusters
from Ellis et al.\ (1997), adopting an arbitrary normalisation.

}

\section{Discussion and Conclusions}

Our aim in this paper is to relate quantitatively the various populations
of galaxies and AGN now being uncovered in large numbers at $z\sim 2$.  
To give a qualitative indication of the form this relation might take we
note that one popular model for the creation of luminous, metal-rich
elliptical galaxies has them formed through a short, intense but highly
obscured, burst of star formation at high redshifts.  Such activity would
be naturally related to the SMG population (Lilly et al.\ 1999; Smail et
al.\ 2002b). Similarly, the proposed presence of supermassive black holes
in all luminous ellipticals and the claimed correlation of their black
hole masses with the stellar mass of their host spheroids, argues that
their progenitors are all likely to have exhibited some form of nuclear
activity during their formation phase. Can we therefore use the space
densities and likely lifetimes of the SMG, QSO and elliptical galaxy
populations to test a simple framework where SMGs represent the monolithic
formation of massive elliptical galaxies and evolve through a QSO phase
(Sanders et al.\ 1988; Page et al.\ 2004)?

The volume density of bright SMGs is $3\times 10^{-5}$\,Mpc$^{-3}$ at
$z\sim 2.5$ (Chapman et al.\ 2003a, 2004a).  Including the submm-faint,
but similarly far-infrared luminous, OFRG population will roughly double
this estimated space density.  The combined population will then be about
$10\times$ more numerous than optically-selected QSOs with $M_B>-25$ at
this epoch (Boyle et al.\ 2000).  Similarly, luminous, evolved galaxies at
$z\sim 1$ have a space density which is about $10\times$ higher than that
of SMGs (Cimatti et al.\ 2002), and comparable to that of $>L^\ast$
elliptical galaxies at $z\sim 0$.  The redshift range covered by the
bright SMG population corresponds to a timespan of roughly 1--2\,Gyrs,
suggesting that if individual SMGs are identified with luminous, evolved
galaxies at $z\sim 1$ and $>L^\ast$ ellipticals at the present-day then
individual SMG's must have lifetimes of only 100--200\,Myrs (see also
Serjeant \& Takagi 2004).  We have shown earlier that this lifetime is
sufficient to allow these galaxies to build up an L$^\ast$'s worth of
stars, given the SFRs inferred from the far-infrared luminosities of
bright SMGs ($\sim 10^3$\,M$_\odot$\,yr$^{-1}$). Indeed, the
reddening-corrected, restframe $V$-band magnitudes of these galaxies imply
that many already have substantial stellar masses (consistent with them
being L$^\ast$ galaxies).  The ratio of SMG to QSO volume densities would
then imply that the latter phase lasts around 10--20\,Myrs.  This is then
broadly consistent with independent estimates of the lifetime of bright
QSOs (Martini \& Weinberg 2001) as well as the relative ratio of
submm-detected QSOs as a fraction of the QSO and SMG populations (Stevens
et al.\ 2004; Chapman et al.\ 2004a).

Thus we have a quantitatively self-consistent scenario. SMG/OFRG's are
cumulative $\sim $\,100--200\,Myr-long star formation events which build
the stellar population of an $\gs$\,L$^\ast$ elliptical galaxy at $z\sim
2$--3, before going through a $\sim $\,10\,Myr-long QSO phase and then
evolving to become the passive EROs seen at $z\sim 1$ (perhaps via a phase
with $(J-K)\geq 2.3$ at $z\sim 2$, Cimatti et al.\ 2004) and subsequently
the luminous elliptical population seen in intermediate and low-redshift
clusters.

The alert reader will notice that there is no mention of the important
UV-selected high-redshift galaxy populations in the framework outlined
above.  This is because we believe there are strong reasons to expect that
the typical UV-selected, star forming galaxy at $z\sim 2$--3 is less
massive than those identified through current far-infrared surveys.  This
belief is based in part upon the differences in the restframe optical
luminosities of the two populations we have uncovered.  It is strengthened
by the apparent differences in the dynamical mass estimates of small
samples from the two populations (Swinbank et al.\ 2004; Erb et al.\ 2003)
and from the strong clustering seen in far-infrared-selected population
(Blain et al.\ 2004a).  All of these observations point toward SMG/OFRG's
being several times more massive than typical UV-selected, high-redshift
galaxies.  Hence, it is unlikely that a typical UV-selected galaxy will
experience a ULIRG-like phase in its evolution -- although mergers of two
such systems may produce sufficiently intense activity.

There are, however, problems with the simple evolutionary picture we have
outlined.  Not least of these are the suggestions that both SMGs (and the
very red near-infrared population at $z\sim 2$) are strongly clustered
with correlation lengths of order $\sim 8 h^{-1}$\,Mpc (Blain et al.\
2004a; Daddi et al.\ 2003).  This is roughly twice that measured for
typical QSOs at this epoch (although perhaps consistent with the
brightest, see Croom et al.\ 2000).  This may indicate that bright
SMG/OFRG's do not evolve into typical QSOs, or that a more complex
temporal bias is operating (Scannapieco \& Thacker 2003). More reliable
clustering measurements will be needed for all populations before this is
a critical issue -- yet it remains a concern.

A second test of this simple evolutionary picture will be provided from
more reliable measurements of the stellar masses of the various
populations.  At $z\sim 2$--3 this requires sensitive mid-infrared imaging
to access the restframe near-infrared emission, where effects from recent
star formation and reddening are minimised.  {\it Spitzer} has
demonstrated the capability to detect the majority of the host galaxies of
high-redshift, far-infrared selected galaxies (Egami et al.\ 2004; Ivison
et al.\ 2004 Frayer et al.\ 2004; Serjeant et al.\ 2004; Charmandaris et
al.\ 2004), and promises to be an important tool for further study of this
population.

We summarise our main conclusions:

\begin{enumerate}

\item We have compiled a large catalog of optical/near-infrared photometry
for far-infrared luminous galaxies selected from deep submm and radio
surveys.  In total we have spectroscopic coverage of all 96 galaxies in
our sample.  Comparing the optical/near-infrared properties of our
optically-faint radio galaxy and radio-selected submm subsamples we find
no significant differences between these two populations.  This supports
the claim that they represent related subsets of the high-redshift
far-infrared luminous galaxy population.

\item We explore the optical--near-infrared and near-infrared colors of
these galaxies and show that they span a large range as a function of
apparent magnitude or redshift.  We attribute this to a wide variation in
obscuration and structure within these systems. We demonstrate that
signatures of dust are discernable in both the high resolution imaging
from {\it HST} of these galaxies and their broad-band photometry.  The
complex obscuration suggested by our observations may be a direct result
of outflows and winds driving channels surrounding dust, providing a
natural explanation of the apparent ease of escape of Ly$\alpha$ photons
from these dusty galaxies (Chapman et al.\ 2004a).

\item We show that typical SMG/OFRG are physically larger than
optically-selected galaxies at similar redshifts drawn from
magnitude-limited samples.  We interpret this as primarily due to the
multi-component nature of the SMG/OFRG, which in turn reflects the central
role of tidal interactions and mergers in triggering the obscured,
luminous activity which is used to select the galaxies in our sample.  
This is confirmed by the clear merger/interacting morphologies shown by a
large fraction of galaxies in our sample within the GOODS HDF-N field,
which were serendipitously imaged by {\it Hubble Space Telescope}.

\item We construct a near-infrared Hubble diagram for our large,
homogeneous sample.  This shows that the $K$-band luminosities of
SMG/OFRGs are typically 1 magnitude fainter than similarly distant
high-redshift, luminous radio galaxies, and show a larger scatter than
radio galaxy samples (Serjeant et al.\ 2003).  We do not find any
statistically-compelling evidence that the most bolometrically-luminous
SMG/OFRGs (those with $>10^{13}L_\odot$) exhibit a smaller dispersion in
their $K$-band magnitudes than the less luminous systems (c.f.\ Serjeant
et al.\ 2003).  This is unsurprising given the multi-component nature of
many of these galaxies and the wide range in their restframe optical
obscuration (Serjeant et al.\ 2003).

\item The optical-near-infrared colors and near-infrared photometry of
SMG/OFRG shows that they are both brighter and redder than either
UV-selected, star-forming galaxies at $z\sim 3$ or $z\sim 2$. We attribute
these differences to larger stellar masses and higher obscuration in the
SMG/OFRG population, resulting from their more massive progenitors and
more active star formation.  Simple photometric modeling appears to
confirm that the restframe SEDs of the SMG/OFRG indicate continuum
reddening at least twice that of the LBG population and their stellar
luminosities may be 5 times larger.

\item We also compare the restframe optical properties of these very
luminous far-infrared galaxies, with similar galaxies in the local
Universe.  We find that these high redshift galaxies extend the trend for
lower optical/far-infrared flux ratios at higher luminosities seen by
previous workers -- suggesting that an increasing fraction of the activity
in these systems is almost completely obscured by dust.  The enhancement
seen in the far-infrared luminosities of the distant population is also
shown by their larger gas masses and higher gas fractions, suggesting a
constant star formation efficiency in the most vigorous starbursts out to
$z\sim 3$.

\item Finally, we show that if we take the star formation properties of a
typical SMG/OFRG from our crude modeling of their restframe UV/optical
SEDs and let them passively evolve from $z\sim 2.2$, they provide a good
fit for the bright-end of the luminosity function of
morphologically-selected spheroidal galaxies in rich clusters at $z\sim
0.5$.  This provides additional support for the claims that SMG/OFRG
represent a highly obscured and very active phase in the early evolution
of massive, elliptical galaxies.

\end{enumerate}

\acknowledgments

We thank Mark Sullivan for obtaining the $J$/$K$-band imaging of the
SA\,22 field, Kevin Bundy, Richard Ellis and Chris Conselice for sharing
their HDF-N $K$ imaging and the UKIRT observers who undertook our queue
observations.  We acknowledge useful conversations or help from Dave
Alexander, Omar Almaini, Carlton Baugh, Colin Borys, Peter Draper, Dave
Frayer, Carlos Frenk, Thomas Greve, Bill Keel, Cedric Lacey, Alice
Shapley, Jason Stevens, Mark Swinbank, Neil Trentham and Chris Willott. We
thank the referee, Dr.\ Steve Serjeant, for a very constructive report
which clarified the discussion and conclusions of this work. IRS
acknowledges support from the Royal Society, AWB acknowledges support from
NSF grant AST-02059377, the Research Corporation and the Alfred P.\ Sloan
Foundation.  We acknowledge use of the 2MASS survey data provided through
IPAC at Caltech.  The Hale 5-m of the Palomar Observatory is owned and
operated by the California Institute of Technology.  UKIRT is operated by
the Joint Astronomy Centre on behalf of the UK Particle Physics and
Astronomy Research Council.  The AAT is operated by the Anglo-Australian
Observatory on behalf of the Australian Research Council and the UK
Particle Physics and Astronomy Research Council.  This research is based
on observations made with the {\it Hubble Space Telescope} retrieved from
the ESO/ST-ECF Science Archive Facility.

%
%
%

%
%
\begin{center}{\small
\begin{table*}[h]
\centerline{\sc Table 1}
\centerline{\sc Photometry of SMGs and OFRGs}
\centerline{\begin{tabular}{cc cccl}
ID & $z$ & $K$ & $J$ & $I$ & Comment\\
\hline\hline
SMM\,J030252.50+000856.4 & 0.176 & 16.15$\pm$0.01 & 17.58$\pm$0.02 & 20.80$\pm$0.03 & CFRS03.10 \\  
SMM\,J030236.15+000817.1 & 2.435 & $\geq$20.9 & $\geq$21.3 & $\geq$23.7 & CFRS03.6 \\ 
SMM\,J030227.73+000653.5 & 1.407 &  19.14$\pm$0.03 & 19.66$\pm$0.04 & 21.29$\pm$0.05 & CFRS03.15 \\  
SMM\,J030231.81+001031.3 & 1.316 & $\geq$20.9 & ...   & $\geq$23.7 & CFRS03.17 \\  
SMM\,J030238.62+001106.3 & 0.276 & 20.52$\pm$0.17 & ...   & 23.14$\pm$0.17 & CFRS03.25 \\ 
SMM\,J105238.30+572435.8 & 3.036 & 20.32$\pm$0.24 & ...   & 23.26$\pm$0.16 & LH\,850.2\\ 
SMM\,J105158.02+571800.3 & 2.239 & 18.86$\pm$0.09 & ...   & 23.24$\pm$0.15 & LH\,850.3 \\ 
SMM\,J105230.73+572209.5$^\ast$ & 2.611 & 19.22$\pm$0.16 & ...   & 22.71$\pm$0.07 & LH\,850.6 \\  
SMM\,J105200.26+572421.7$^\ast$ & 0.689 & 18.82$\pm$0.09 & ...   & 21.59$\pm$0.03 & LH\,850.8 \\  
SMM\,J105207.49+571904.0$^\ast$ & 2.694 & $\geq$20.6 & ...  & 22.66$\pm$0.08 & LH\,850.12 \\  
... & ... & ... & ... & ... & ... \\
... & ... & ... & ... & ... & ... \\
\hline
\end{tabular}}
\smallskip
{\small $\ast$) Photometry may be contaminated by near neighbor. }
\end{table*}}
\end{center}


\begin{thebibliography}{}

\bibitem[]{} Baugh, C.M., Lacey, C.G., Frenk, C.S., Granato, G.L.,
Silva, L., Bressan, A., Benson, A.J., Cole, S., 2004, MNRAS, submitted.

\bibitem[]{1227} Bergstrom, S, Wiklind, T., 2004, A\&A, 414, 95

\bibitem[]{1229} Bertoldi, F., Carilli, C.L., Menten, K.M., Owen, F., Dey, A.,
Gueth, F., Graham, J.R., Kreysa, E., et al., 2000, A\&A, 360, 92

\bibitem[]{1232} Blain, A.W., Chapman, S.C., Smail, I., Ivison, R.J., 2004a,
ApJ, in press

\bibitem[]{1235} Blain, A.W., Chapman, S.C., Smail, I., Ivison, R.J., 2004b,
ApJ, in press

\bibitem[]{1238} Bolzonella, M., Miralles, J.-M., Pello, R.,
 2000, A\&A, 363, 476

\bibitem[]{1241} Borys, C., et al., 2004, MNRAS, submitted

\bibitem[]{1243} Boyle, B.J., Shanks, T., Croom, S.M., Smith R.J., Miller, L.,
Loaring, N., Heymans, C., 2000, MNRAS, 317, 1014

\bibitem[]{1246} Bundy, K., et al., 2004, in prep

\bibitem[]{1248} Capak, P., Cowie, L.L., Hu, E.M., Barger, A.J., Dickinson,
M., Fernandez, E., Giavalisco, M., Komiyama, Y., et al., 2003, AJ, 127,
180

\bibitem[]{1252} Chapman, S.C., 
Scott, D., Steidel, C.C., Borys, C., Halpern, M., Morris, S.L., 
Adelberger, K.L., Dickinson, M., et al.,
2000, MNRAS, 319, 318 

\bibitem[]{1257} Chapman, S.C., Blain, A.W., Ivison, R.J., Smail, I., 2003a,
Nature, 422, 695

\bibitem[]{1260} Chapman, S.C., Windhorst, R., Odewahn, S., Yan, H.,
Conselice, C., 2003b, ApJ, 599, 92

\bibitem[]{1263} Chapman, S.C., Blain, A.W., Smail, I., Ivison, R.J., 2004a,
ApJ, submitted

\bibitem[]{1266} Chapman, S.C., Smail, I., Blain, A.W., Ivison, R.J., 2004b,
ApJ, in press

\bibitem[]{} Charmandaris, V., et al., 2004,  ApJS, in press

\bibitem[]{1269} Clements, D.L., Eales, S.A., Wojciechowski, K., Webb, T.,
Lilly, S.J., Dunne, L., Ivison, R.J., McCracken, H., et al., 2004, MNRAS,
in press

\bibitem[]{1273} Cimatti, A., et al., 2002, A\&A, 381, L68

\bibitem[]{} Cimatti, A., et al., 2004, Nature, in press

\bibitem[]{1275} Cole, S.C., Norberg, P., Baugh, C.M., Frenk, C.S.,
Bland-Hawthorn, J., Bridges, T., Cannon, R., Colless, M., et al., 2001,
MNRAS, 326, 255

\bibitem[]{1279} Croom, S.M., Boyle, B.J., Loaring, N.S., Miller, L., Outram, P.J., 
Shanks, T., Smith, R.J., 2002, MNRAS, 335, 459

\bibitem[]{1282} Daddi, E., Rottgering, H.J.A., Labbe, I., Rudnick, G., Franx,
M., Moorwood, A.F.M., Rix, H.W., van der Werf, P.P., van Dokkum, P.G.,
2003, 588, 50

\bibitem[]{1286} Daddi, E., Cimatti, A., Renzini, A., Vernet, J., Conselice,
C., Pozzetti, L., Mignoli, M., Tozzi, P., et al., 2004, 600, L127

\bibitem[]{1289} Dannerbauer, H., Lehnert, M.D., Lutz, D., Tacconi, L.,
Bertoldi, F., Carilli, C., Genzel, R., Menten, K.M., 2004, A\&A, submitted

\bibitem[]{1292} Dressler, A., Oemler, A., Couch, W.J., Smail, I., Ellis,
R.S., Barger, A., Butcher, H., Poggianti, B.M., Sharples, R.M., 1997, ApJ,
490, 577

\bibitem[]{} Egami, E., et al., 2004,   ApJS, in press

\bibitem[]{1296} Eikenberry et al., 2004, WIRC2 Manual, Palomar Observatory

\bibitem[]{1298} Ellis, R.S., Smail, I., Dressler, A., Couch, W.J., Oemler,
A., Butcher, H., Sharples, R.M., 1997, ApJ, 483, 582

\bibitem[]{1301} Erb, D.K., Shapley, A.E., Steidel, C.C., Pettini, M.,
Adelberger, K.L., Hunt, M.P., Moorwood, A.F.M., Cuby, J., 2003, ApJ, 591, 101

\bibitem[]{1304} Ferguson, H.C., Dickinson, M., Giavalisco, M., Kretchmer, C.,
Ravindranath, S., Idzi, R., Taylor, E., Conselice, C.J., et al., 2004,
ApJ, 600, L107

\bibitem[]{1308} Fioc, M., Rocca-Volmerange, B., 1997, A\&A, 326, 950

\bibitem[]{1310} Franx, M., Labbe, I., Rudnick, G., van Dokkum, P.G., Daddi,
E., Forster Schreiber, N.M., Moorwood, A., Rix, H.W., et al., 2003, 587,
L79

\bibitem[]{1314} Frayer, D.T., Ivison, R.J., Scoville, N.Z., Yun, M., Evans,
A.S., Smail, I., Blain, A.W., Kneib, J.-P., 1998, ApJL, 506, L7

\bibitem[]{1317} Frayer, D.T., Ivison, R.J., Scoville, N.Z., Evans, A.S., Yun,
M., Smail, I., Barger, A.J., Blain, A.W., Kneib, J.-P., 1999, ApJL, 514,
L13

\bibitem[]{1321} Frayer, D.T., Reddy, N.A, Armus, L., Blain, A.W., Scoville,
N.Z., Smail, I., 2004, AJ, 127, 728

\bibitem[]{} Frayer, D.T., et al.,  2004,  ApJS, in press

\bibitem[]{1324} Gear, W.K., Lilly, S.J., Stevens, J.A., Clements, D.L., Webb,
T.M.A., Eales, S.A., Dunne, L., 2000, MNRAS, 316, L51

\bibitem[]{1327} Genzel, R., Baker, A.J., Tacconi, L.J., Lutz, D., Cox, P.,
Guilloteau, S., Omont, A., 2003, ApJ, 584, 633

\bibitem[]{1330} Giavalisco, M., Ferguson, H.C., Koekemoer, A.M., Dickinson,
M., Alexander, D.M., Bauer, F.E., Bergeron, J., Biagetti, C., 2004, ApJ,
600, L93

\bibitem[]{1334} Greve, T., et al., 2004a, in prep

\bibitem[]{1336} Greve, T., et al., 2004b, MNRAS, in press

\bibitem[]{1338} Ivison, R.J., Greve, T.R., Smail, I., Dunlop, J.S., Roche,
N.D., Scott, S.E., et al., 2002, MNRAS, 337, 1

\bibitem[]{1341} Ivison, R.J., et al., 2004, ApJS, in press

\bibitem[]{1343} Kim, D.-C., Veilleux, S., Sanders, D.B., 2002, ApJS, 143, 277

\bibitem[]{1345} Kochanek, C., Pahre, M.A., Falco, E.E., Huchra, J.P., Mader,
J., Jarrett, T.H., Chester, T., Cutri, R., Schneider, S.E., 2001, ApJ,
560, 566

\bibitem[]{1349} Kodama, T., Yamada, T., Akiyama, M., Aoki, K., Doi, M.,
Furusawa, H., Fuse, T., Imanishi, M., et al., 2004, MNRAS, in press

\bibitem[]{1352} Landolt, A.U., 1992, AJ, 104, 340

\bibitem[]{1354} Leitherer, C., Schaerer, D., Goldader, J.D., Delgado, R.M.G., Robert, 
C., Kune, D.F., de Mello, D.F., Devost, D., Heckman, T.M., 1999, ApJS,

\bibitem[]{1357} Lilly, S.J., et al., 1999, ApJ, 518, 614

\bibitem[]{1359} de Lucia, G., Poggianti, B.M., Aragon-Salamanca, A., Clowe,
D., Halliday, C., Jablonka, P., Milvang-Jensen, B., Pello, R., et al.,
2004, ApJ, submitted

\bibitem[]{1363} Lutz, D., Dunlop, J.S., Almaini, O., Andreani, P., Blain,
A.W., Efstathiou, A., Fox, M., Genzel, R., et al., 2001, A\&A, 378, L70

\bibitem[]{1366} Mihos, J.C., Hernquist, L., 1996, ApJ, 464, 641

\bibitem[]{1368} Neri, R., Genzel, R., Ivison, R.J., Bertoldi, F.,
Blain, A.W., Chapman, S.C., Cox, P., Greve, T.R., Omont, A., Frayer,
D.T., 2003, ApJ, 597, L113 

\bibitem[]{} Page, M.J., Stevens, J.A., Ivison, R.J., Carrera, F.J., 
2004, ApJ, submitted
 
\bibitem[]{} Pentericci, L., et al., 2001, ApJS, 135, 63

\bibitem[]{1372} Pozzetti, L., Mannucci, F., 2000, MNRAS, 317, L17

\bibitem[]{1374} Sanders, D.B., Soifer, B.T., Elias, J.H., Neugebauer, G.,
Matthews, K., 1988, ApJ, 328, L35

\bibitem[]{1377} Scannapieco, E., Thacker, R.J., 2003, ApJ, 590, L69

\bibitem[]{1379} Scoville, N.Z., et al., 2000, AJ, 119, 991 

\bibitem[]{1381} Serjeant, S., Farrah, D., Geach, J., Takagi, T., Verma, A.,
Kaviani, A., Fox, M., 2003, MNRAS, 346, L51

\bibitem[]{} Serjeant, S., Takagi, T., 2004, Nature, submitted

\bibitem[]{} Serjeant, S., et al., 2004, ApJS, in press

\bibitem[]{1384} Shapley, A.E., Steidel, C.C., Adelberger, K.L., Dickinson,
M., Giavalisco, M., Pettini, M., 2001, ApJ, 562, 95
 
\bibitem[]{1387} Smail, I., Ivison, R.J., Blain, A.W., Kneib, J.-P., 
1998, ApJL, 507, L21

\bibitem[]{1390} Smail, I., Ivison, R.J., Kneib, J.-P., Cowie, L.L., Blain,
A.W., Barger, A.J., Owen, F.N., Morrison, G.E., 1999a, MNRAS, 308, 1061

\bibitem[]{1393} Smail, I., Morrison, G., Gray, M.E., Owen, F.N., Ivison,
R.J., Kneib, J.-P., Ellis, R.S., 1999b, ApJ, 525, 609

\bibitem[]{1396} Smail, I., Owen, F.N., Morrison, G.E., Keel, W.C., Ivison,
R.J., Ledlow, M.J., 2002a, ApJ, 581, 844

\bibitem[]{1399} Smail, I., Ivison, R.J., Blain, A.W., Kneib, J.-P., 2002b,
MNRAS, 331, 495

\bibitem[]{1402}
Smail, I., Chapman, S.C., Ivison, R.J., Blain, A.W., Takata, T., Heckman, T.M., 
Dunlop, J.S., Sekiguchi, K., 2003a, MNRAS, 342, 1185

\bibitem[]{1406} Smail, I., Scharf, C.A., Ivison, R.J., Stevens, J.A., 
Bower, R.G., Dunlop, J.S., 2003b, ApJ, 599, 86

\bibitem[]{1409} Stanford, S.A., Stern, D., van Breugel, W., de Breuck, C.,
2000, ApJS, 131, 185

\bibitem[]{1412} Steidel, C.C., Shapley, A.E., Pettini, M., Adelberger, K.L.,
Erb, D.K., Reddy, N.A., Hunt, M.P., 2004, ApJ, 604, 534

\bibitem[]{} Stevens, J.A., Page, M.J., Ivison, R.J., Mittaz, J.P.D.,
Carrera, F.J., Smail, I., McHardy, I.M., 2004, MNRAS, submitted

\bibitem[]{1415} Swinbank, A.M., et al., 2004, ApJ, submitted

\bibitem[]{1417} Tinney, C., et al., 2004, IRIS2 manual, AAO.

\bibitem[]{1419} Trentham, N., Kormendy, J., Sanders, D.B., 1999, 117, 2152

\bibitem[]{1421} Veilleux, S., Kim, D.-C., Sanders, D.B., Mazzarella, J.M.,
Soifer, B.T., 1995, ApJS, 98, 171

\bibitem[]{1424} Webb, T.M.A., Eales, S.A., Lilly, S.J., Clements, D.L.,
Dunne, L. Gear, W.K., Ivison, R.J., Flores, H., Yun, M., 2003, ApJ, 587,
41

\bibitem[]{1428} Webb, T.M.A., Brodwin, M., Eales, S.A., Lilly, S.J., 2004,
ApJ, in press

\bibitem[]{1431} Martini, P., Weinberg, D.H., 2001, ApJ, 547, 12

\bibitem[]{1433} Willott, C., Rawlings, S., Jarvis, M.J., Blundell, K.M.,
2003, MNRAS, 339, 173

\bibitem[]{} Yagi, M., 1998, PhD Thesis, University of Tokyo

\end{thebibliography}
\end{document}